%February 8, 2004
\input amstex
\magnification 1200
\TagsOnRight
\def\qed{\ifhmode\unskip\nobreak\fi\ifmmode\ifinner\else
 \hskip5pt\fi\fi\hbox{\hskip5pt\vrule width4pt
 height6pt depth1.5pt\hskip1pt}}
\NoBlackBoxes
%\hyphenation{pre-print}
\baselineskip 18 pt
\parskip 5 pt
\def\stretch {\noalign{\medskip}}
\define \bCp {\bold C^+}
\define \bCm {\bold C^-}
\define \bCpb {\overline{\bold C^+}}
\define \ds {\displaystyle}
\define \bR {\bold R}
\define \bm {\bmatrix}
\define \endbm {\endbmatrix}

\centerline {\bf INVERSE SPECTRAL-SCATTERING PROBLEM}
\centerline {\bf WITH TWO SETS OF DISCRETE SPECTRA}
\centerline {\bf FOR THE RADIAL SCHR\"ODINGER EQUATION}

\vskip 10 pt
\centerline {Tuncay Aktosun}
\vskip -8 pt
\centerline {Department of Mathematics and Statistics}
\vskip -8 pt
\centerline {Mississippi State University}
\vskip -8 pt
\centerline {Mississippi State, MS 39762, USA}
\vskip -8 pt
\centerline {aktosun\@math.msstate.edu}

\centerline {Ricardo Weder\plainfootnote{$^\dagger$}
{Fellow Sistema Nacional de Investigadores}}
\vskip -8 pt
\centerline {Instituto de Investigaciones en
Matem\'aticas Aplicadas y en Sistemas}
\vskip -8 pt
\centerline {Universidad Nacional Aut\'onoma de M\'exico}
\vskip -8 pt
\centerline {Apartado Postal 20-726, IIMAS-UNAM, M\'exico DF 01000, M\'exico}
\vskip -8 pt \centerline {weder\@servidor.unam.mx}

\noindent {\bf Abstract}: The Schr\"odinger equation on the half
line is considered with a real-valued, integrable potential having
a finite first moment. It is shown that the potential and the
boundary conditions are uniquely determined by the data
containing the discrete
eigenvalues for a boundary condition
at the origin, the continuous part of the spectral measure
for that boundary condition, and a
subset of the discrete eigenvalues for a different boundary
condition. This result extends the celebrated two-spectrum
uniqueness theorem of Borg and Marchenko to the case where there
is also a continuous spectrum.

\vskip -8 pt
%\par\noindent {\bf Short title:} Inverse radial scattering
\vskip -8 pt
%\par\noindent {\bf Alternate title:} Generalized Borg-Marchenko theory with continuous spectrum
\newpage

\noindent {\bf 1. INTRODUCTION}
\vskip 3 pt

The inverse spectral theory deals with the determination of a
differential operator from an appropriate set of spectral data.
Its origin goes back to Ambartsumyan [1] who considered the
Sturm-Liouville problem
$$-\psi''+V(x)\,\psi=\lambda\,\psi,\qquad x\in (0,\pi),\tag 1.1$$
$$\psi'(0)=\psi'(\pi)=0,$$
where the prime denotes the spatial $x$-derivative and the
potential $V$ is continuous and real valued. Ambartsumyan
indicated that if $\{\lambda_j\}_{j=0}^\infty$ is the eigenvalue
set for this Sturm-Liouville problem and if $\lambda_j=j^2$ for
$j=0,1,2,\dots,$ then $V\equiv 0.$ Next, Borg showed [2] that one
spectrum in general does not uniquely determine the corresponding
Sturm-Liouville operator and that Ambartsumyan's result is really
a special case. In particular, Borg gave the proof of the
following result: Let $\{\lambda_j\}_{j=0}^\infty$ be the
eigenvalue set for (1.1) with the boundary conditions
$$\cos\alpha\cdot \psi'(0)+\sin\alpha\cdot\psi(0)=0,\quad
\cos\beta\cdot\psi'(\pi)+\sin\beta\cdot\psi(\pi)=0,$$
and let $\{\mu_j\}_{j=0}^\infty$ be the eigenvalue set
with the boundary condition
$$\cos\gamma\cdot\psi'(0)+\sin\gamma\cdot\psi(0)=0,\quad
\cos\beta\cdot\psi'(\pi)+\sin\beta\cdot\psi(\pi)=0,$$
where $\gamma\neq\alpha.$ Then, the two sets
$\{\lambda_j\}_{j=0}^\infty$ and $\{\mu_j\}_{j=0}^\infty$ uniquely
determine $\alpha,$ $\beta,$ $\gamma,$ and $V.$

Borg [3] and Marchenko [4] studied the Sturm-Liouville operator on
the half line with a boundary condition at the origin when there
is no continuous spectrum. They showed that two sets of discrete
spectra associated with distinct boundary conditions at $x=0$
(with a fixed boundary condition, if any, at $x=+\infty$)
uniquely determine the potential and the boundary conditions at
the origin.

A continuous spectrum arises in applications often. It comes into
play in a natural way in the analysis of potentials vanishing at
infinity. In this paper we generalize the celebrated
Borg-Marchenko result to the case where there is also a continuous
spectrum; namely, we prove that the potential and boundary
conditions are uniquely determined by an appropriate data set
containing the discrete eigenvalues and continuous part of the
spectral measure corresponding to one boundary condition at the
origin and a subset of the discrete eigenvalues for a different
boundary condition. Another extension of the Borg-Marchenko
theorem to the case with a continuous spectrum is given by
Gesztesy and Simon [5], where a uniqueness result is presented
when Krein's spectral shift function is known. In our
generalization of the Borg-Marchenko theorem, our conditions are
directly stated in terms of the spectral measure; namely, the
amplitude of the Jost function and the eigenvalues. There is an
extensive literature on the inverse spectral problem; for other
important contributions to the field and a more detailed
historical account, we refer the reader to [5-8].

Consider the radial Schr\"odinger equation, related to (1.1) with $\lambda=k^2,$
$$-\psi''+V(x)\,\psi=k^2\psi,\qquad x\in(0,+\infty),\tag 1.2$$
with the boundary condition
$$\sin\alpha\cdot \psi'(k,0)+\cos\alpha\cdot\psi(k,0)=0,\tag 1.3$$
for some $\alpha\in(0,\pi].$ The condition (1.3) is also written
as
$$\cases \psi'(k,0)+\cot\alpha\cdot\psi(k,0)=0,\qquad \alpha\in(0,\pi),\\
\stretch
\psi(k,0)=0,\qquad \alpha=\pi.\endcases\tag 1.4$$
In (1.3) or (1.4) we get
the Dirichlet condition if $\alpha=\pi,$ the Neumann condition
if $\alpha=\pi/2,$ and otherwise the mixed condition. We assume
that the potential $V$ in (1.2) belongs to the Faddeev class;
i.e., it is real valued and belongs to $L^1_1(\bR^+),$ where
$L^1_n(J)$ denotes the Lebesgue-measurable functions $V$ defined
on a Lebesgue-measurable set $J$ for which $\int_J
dx\,(1+|x|)^n|V(x)|$ is finite.

Let $H_\alpha$ for $\alpha\in(0,\pi]$ denote the unique selfadjoint realization
[9] of $-d^2/dx^2+V$ in $L^2(0,+\infty)$ with the boundary
condition (1.3). It is known [8,9] that $H_\alpha$ has no
positive or zero eigenvalues, it has no singular-continuous
spectrum, and its absolutely-continuous spectrum consists of
$[0,+\infty).$ It has a finite number of simple negative
eigenvalues, and we use $\sigma_{\text d}(H_\alpha):=
\{-\kappa_{\alpha j}^2\}_{j=1}^{N_\alpha}$ to denote
the eigenvalue set. The Jost function of (1.2) associated with the
boundary condition (1.4) is defined as [8]
$$F_\alpha(k):=\cases -i[f'(k,0)+\cot\alpha\cdot f(k,0)],\qquad
\alpha\in(0,\pi),\\
\stretch
f(k,0),\qquad \alpha=\pi,\endcases\tag 1.5$$
where $f(k,x)$
denotes the Jost solution to (1.2) satisfying the asymptotics
$$f(k,x)=e^{ikx}[1+o(1)],\quad
f'(k,x)=ik\,e^{ikx}[1+o(1)],\qquad x\to+\infty.\tag 1.6$$
It is known [7,8] that the set $\{i\kappa_{\alpha j}\}_{j=1}^{N_\alpha}$
corresponds to the zeros of $F_\alpha$ in
$\bCp.$ We use $\bCp$ for the upper half complex plane and
$\bCpb:=\bCp\cup\bR$ for its closure.

There are two main methods to solve the inverse spectral and
scattering problems for the radial Schr\"odinger equation; namely,
the Gel'fand-Levitan method and the Marchenko method. The former
[7,8,10,11] solves the inverse spectral problem, and the potential
and boundary condition are uniquely reconstructed by solving the
Gel'fand-Levitan integral equation (5.4) with the input data (5.5) or (5.6)
obtained from the spectral measure $\rho_\alpha(\lambda)$
given in (3.23). The part of the spectral
measure associated with the continuous spectrum is absolutely
continuous, and as seen from (3.23) its derivative
at energy $k^2$ is determined by $|F_\alpha|.$ The part
associated with the discrete spectrum is determined by the
set of eigenvalues $\{-\kappa_{\alpha j}^2\}_{j=1}^{N_\alpha}$ and the norming
constants $\{g_{\alpha j}\}_{j=1}^{N_\alpha}.$ On the other hand,
the Marchenko method [7,8,10,12] is an inverse scattering procedure,
and the potential and boundary condition are uniquely
reconstructed by solving the Marchenko integral equation (5.10) in
terms of the scattering data consisting [cf. (5.7) and (5.11)] of
the scattering matrix $S_\alpha,$ the bound state energies
$\{-\kappa_{\alpha j}^2\}_{j=1}^{N_\alpha},$ and the
norming constants $\{m_{\alpha j}\}_{j=1}^{N_\alpha},$
where the scattering matrix is defined as
$$S_\alpha(k):=\cases -\ds\frac{F_\alpha(-k)}{F_\alpha(k)},
\qquad \alpha\in(0,\pi),\\
\stretch
\ds\frac{F_\pi(-k)}{F_\pi(k)},
\qquad \alpha=\pi.\endcases\tag 1.7$$

Our generalized
Borg-Marchenko problem is stated as follows.
Let $\beta\in(0,\pi)$ with $\beta<\alpha\le \pi$
correspond to the boundary condition obtained from (1.3) by
replacing $\alpha$ there with $\beta.$ This leads to, via (1.5),
the Jost function $F_\beta$ with zeros at $k=i\kappa_{\beta j}$
in $\bCp,$ where $j=1,\dots,N_\beta.$  Assume that we are
given some data $\Cal D,$ which contains $|F_\alpha|$ for $k\in\bR,$
the whole set $\{\kappa_{\alpha j}\}_{j=1}^{N_\alpha},$ and a
subset of $\{\kappa_{\beta j}\}_{j=1}^{N_\beta}$ consisting of
$N_\alpha$ elements. Alternatively, our data $\Cal D$ may include
$|F_\beta(k)|$ for $k\in\bR$ and the sets
$\{\kappa_{\alpha j}\}_{j=1}^{N_\alpha}$ and
$\{\kappa_{\beta j}\}_{j=1}^{N_\beta}.$
Does $\Cal D$ uniquely determine the set
$\Cal E,$ where $\Cal E:=\{V,\alpha,\beta\}$? If not, what
additional information do we need besides $\Cal D$ in order to
determine $\Cal E$ uniquely? Can we present a constructive method to recover
$\Cal E$ from $\Cal D$ or from a data set
obtained by some augmentation of $\Cal D$?

This generalized Borg-Marchenko problem can be considered as an
inverse scattering problem because both the Faddeev class of
potentials and the Jost function are natural elements in
scattering theory. On the other hand, this problem is also an
inverse spectral problem because in our data we use
$|F_\alpha|$ and $\{-\kappa_{\alpha j}^2\}_{j=1}^{N_\alpha},$
which are both contained in the relevant spectral measure. In
fact, from this point of view, we replace the $N_\alpha$ norming
constants appearing in the discrete portion of the spectral
measure by $N_\alpha$ of the eigenvalues for a different boundary
condition. This constitutes a natural mathematical problem which
is actually an inverse problem with two discrete spectra in the
presence of a continuous spectrum. Replacing the norming constants
in the Gel'fand-Levitan or the Marchenko method by a set of
eigenvalues from a second boundary condition is also interesting from
the viewpoint of physical applications. This is because
eigenvalues have a direct physical interpretation as energies of
the stationary states of a quantum mechanical system, whereas, a
priori, norming constants do not have such a clear physical
interpretation.

Our problem can also be considered as an inverse scattering
problem on the line with a potential supported on a half line. As
we show in Section~5, from our data we can
uniquely construct the data set $\Cal F$ given in
(5.14), which contains enough
information [13-18] to
reconstruct the potential by using any one of
the full-line inversion methods [7,10,19-22].

Our motivation for this paper came from a question by Roy Pike [23]
as whether $f'(k,0):=i\,F_{\pi/2}(k),$ the spatial derivative of
the Jost solution to the one-dimensional Schr\"odinger
equation evaluated at
$x=0,$ can uniquely determine the corresponding potential if
that potential is known to be zero
on the negative half line. This question
arises in the acoustical analysis of the human vocal tract. When
the vocal tract is stimulated by a sinusoidal input volume
velocity at the glottis, the impulse response at the lips is
(cf. (70) in [24]) essentially given by $f'(k,0).$ Such
an inverse problem is equivalent to determining a scaled curvature
of the duct of the vocal tract when a constant-frequency sound is
uttered, and it has important applications in speech recognition [24].

The method we use is a generalization of that used in [23] in
the case of a potential that has no bound states for either the
Dirichlet and Neumann boundary conditions and that is perturbed by
a Dirac delta distribution at $x=0.$ The basic idea is to relate our data
to the real part of a function that is in the Hardy class of
functions analytic in $\bCp.$ It turns out that the real part of
such a function is determined for $k\in\bR$ by our data. Then, the
function itself is uniquely constructed in $\bCpb$ with the help
of the Schwarz integral formula [25-27].
Our proofs also present a method for the reconstruction
of the potential and boundary conditions.

Our paper is organized as follows. We list our main results
as Theorems~2.1-2.8 in Section~2. Then, in
Section~3 we presents the results needed in order
to prepare the proofs of these theorems. In Section~4 the proof
of each theorem is given by a constructive
method; from the appropriate scattering-spectral data sets
$\Cal D_1,\dots,\Cal D_8$ given in (2.4)-(2.11), respectively,
it is shown how the boundary conditions
are uniquely reconstructed and how appropriate information
can be assembled in order to uniquely
reconstruct the potential.
In Section~5 we outline several methods to uniquely
reconstruct the potentials.
Finally, in Section~6, we illustrate the uniqueness and
reconstruction by
some explicit examples.

\vskip 10 pt
\noindent {\bf 2. MAIN THEOREMS}
\vskip 3 pt

In Theorems~2.1-2.8 given below we generalize
the celebrated two-spectra uniqueness theorem proved by Borg [3]
and Marchenko [4] from the case of purely discrete spectra to
the case where there is also a continuous spectrum.
We take into consideration all possibilities with
$N_\alpha=N_\beta$ or $N_\alpha=N_\beta-1,$
with $\alpha\in(0,\pi)$ or $\alpha=\pi,$ and by
using $|F_\alpha|$ or $|F_\beta|$ in our data.

In order to state our results in a
precise way, we introduce some notations. Define
$$h_{\beta\alpha}:=\cot\beta-\cot\alpha,\qquad \alpha,\beta\in(0,\pi).\tag 2.1$$
 From (1.5) and (2.1), for $\alpha\ne \beta$ we get
$$f(k,0)=\cases\ds\frac{i}{h_{\beta\alpha}}\left[F_\beta(k)-F_\alpha(k)\right],
\qquad \alpha,\beta\in(0,\pi),\\
\stretch
F_\pi(k),\endcases\tag 2.2$$
$$f'(k,0)=\cases \ds\frac{i}{h_{\beta\alpha}}\left[\cot\beta\cdot
F_\alpha(k)-\cot\alpha\cdot F_\beta(k)\right],\qquad \alpha,\beta\in(0,\pi),\\
\stretch
i\,F_\beta(k)-\cot\beta\cdot F_\pi(k),\qquad\beta\in(0,\pi).\endcases\tag 2.3$$
Notice that $h_{\beta\alpha}>0$ if $0<\beta<\alpha<\pi$ because
the cotangent function is monotone decreasing on $(0,\pi).$
Let $\tilde V$ be another potential in the
Faddeev class, $\tilde H_{\gamma}$ be the corresponding
realization of $-d^2/dx^2+\tilde V$ in $L^2(0,+\infty)$ with the
boundary condition (1.3) in which $\alpha$ is replaced by $\gamma,$
and $\sigma_{\text d}(\tilde H_{\gamma})$ denote
the corresponding eigenvalue set
$\{-\tilde\kappa_{\gamma j}^2\}_{j=1}^{\tilde N_{\gamma}}.$

Let us introduce the appropriate data sets $\Cal D_1,\dots,\Cal D_8$
used as inputs in Theorems~2.1-2.8, respectively, as follows:
$$\Cal D_1:=\{h_{\beta\alpha},|F_\alpha(k)| \text{ for } k\in\bR,
\{\kappa_{\alpha j}\}_{j=1}^{N_\alpha},
\{\kappa_{\beta j}\}_{j=1}^{N_\beta}\},\tag 2.4$$
$$\Cal D_2:=\{\beta,|F_\pi(k)| \text{ for } k\in\bR,
\{\kappa_{\pi j}\}_{j=1}^{N_\pi},
\{\kappa_{\beta j}\}_{j=1}^{N_\beta}\},\tag 2.5$$
$$\Cal D_3:=\{h_{\beta\alpha},|F_\alpha(k)| \text{ for } k\in\bR,
\{\kappa_{\alpha j}\}_{j=1}^{N_\alpha}, {\text{an $N_\alpha$-element subset of}}
\ \{\kappa_{\beta j}\}_{j=1}^{N_\beta}\},\tag 2.6$$
$$\Cal D_4:=\{\beta,|F_\pi(k)| \text{ for } k\in\bR,
\{\kappa_{\pi j}\}_{j=1}^{N_\pi}, {\text{an $N_\pi$-element subset of}}
\ \{\kappa_{\beta j}\}_{j=1}^{N_\beta}\},\tag 2.7$$
$$\Cal D_5:=\{h_{\beta\alpha},|F_\beta(k)| \text{ for } k\in\bR,
\{\kappa_{\alpha j}\}_{j=1}^{N_\alpha}, \{\kappa_{\beta
j}\}_{j=1}^{N_\beta}\},\tag 2.8$$
$$\Cal D_6:=\{|F_\beta(k)| \text{ for } k\in\bR,
\{\kappa_{\pi j}\}_{j=1}^{N_\pi}, \{\kappa_{\beta
j}\}_{j=1}^{N_\beta}\},\tag 2.9$$
$$\Cal D_7:=\{\beta,h_{\beta\alpha},|F_\beta(k)| \text{ for } k\in\bR,
\{\kappa_{\alpha j}\}_{j=1}^{N_\alpha}, \{\kappa_{\beta
j}\}_{j=1}^{N_\beta}\},\tag 2.10$$
$$\Cal D_8:=\{\beta,|F_\beta(k)| \text{ for } k\in\bR,
\{\kappa_{\pi j}\}_{j=1}^{N_\pi}, \{\kappa_{\beta
j}\}_{j=1}^{N_\beta}\}.\tag 2.11$$

\noindent {\bf Theorem 2.1} {\it Let the realizations $H_\alpha$ and $H_\beta$
correspond to a potential $V$ in the Faddeev class
with the boundary conditions
identified by $\alpha$ and $\beta,$ respectively. Similarly, let $\tilde
H_{\gamma}$ and $\tilde H_\epsilon$ correspond to $\tilde V$
in the Faddeev class with
the boundary conditions identified with $\gamma$ and $\epsilon,$
respectively. Denote the corresponding
Jost functions by $F_\alpha,$ $F_\beta,$ $\tilde F_\gamma,$ and $\tilde F_\epsilon,$
respectively. Suppose that}
\item{(i)} $0<\beta<\alpha<\pi.$
\item{(ii)} $N_\alpha=N_{\beta}\ge 0.$
\item{(iii)} $h_{\beta\alpha}=h_{\epsilon\gamma}.$
\item{(iv)} $\sigma_{\text d}(H_{\alpha})=\sigma_{\text d}(\tilde H_{\gamma}).$
\item{(v)} $\sigma_{\text d}(H_\beta)=\sigma_{\text d}(\tilde H_\epsilon).$
\item{(vi)} $|F_\alpha(k)|=|\tilde F_{\gamma}(k)|$ {\it for} $k\in\bR.$

\noindent {\it Then, we have $\alpha=\gamma,$ $\beta=\epsilon,$
and $V=\tilde V.$ This is equivalent to saying that
if $N_\alpha=N_{\beta}\ge 0$ and $0<\beta<\alpha<\pi$
then the data set $\Cal D_1$ given in (2.4) uniquely determines
$\{V,\alpha,\beta\}.$}

Next, we consider the analog of Theorem~2.1 when $\alpha=\pi.$

\noindent {\bf Theorem 2.2} {\it With the same notation as in
Theorem~2.1, assume that}
\item{(i)} $0<\beta<\alpha=\pi.$
\item{(ii)} $N_\alpha=N_{\beta}\ge 0.$
\item{(iii)} $\beta=\epsilon.$
\item{(iv)} $\sigma_{\text d}(H_{\alpha})=\sigma_{\text d}(\tilde H_{\gamma}).$
\item{(v)} $\sigma_{\text d}(H_\beta)=\sigma_{\text d}(\tilde H_\epsilon).$
\item{(vi)} $|F_\alpha(k)|=|\tilde F_{\gamma}(k)|$ {\it for} $k\in\bR.$

\noindent {\it Then, we have $\alpha=\gamma$ and $V=\tilde V.$
Equivalently,
if $N_\alpha=N_{\beta}\ge 0$ and $0<\beta<\alpha=\pi$
then the data $\Cal D_2$ given in (2.5) uniquely determines
$V.$}

In the next result, the analog of Theorem~2.1 is
considered when
$N_\alpha=N_\beta-1.$

\noindent {\bf Theorem~2.3}
{\it With the same notation as in
Theorem~2.1, suppose that}
\item{(i)} $0<\beta<\alpha<\pi.$
\item{(ii)} $N_\alpha=N_{\beta}-1\ge 0.$
\item{(iii)} $h_{\beta\alpha}=h_{\epsilon\gamma}.$
\item{(iv)} $\sigma_{\text d}(H_{\alpha})=\sigma_{\text d}(\tilde H_{\gamma}).$
\item{(v)} {\it The intersection of $\sigma_{\text d}(H_\beta)$ and
$\sigma_{\text d}(\tilde H_\epsilon)$ contains at least $N_\alpha$ common
elements.}
\item{(vi)} $|F_\alpha(k)|=|\tilde F_{\gamma}(k)|$ {\it for} $k\in\bR.$

\noindent {\it Then, we have $\alpha=\gamma,$ $\beta=\epsilon,$
and $V=\tilde V.$
This is equivalent to saying that
if $N_\alpha=N_{\beta}-1\ge 0$ and $0<\beta<\alpha<\pi$
then $\{V,\alpha,\beta\}$ is uniquely determined by
the data $\Cal D_3$ defined in (2.6).}

In the next theorem we consider
the analog of Theorem~2.3 when
$\alpha=\pi,$ or equivalently, the analog of
Theorem~2.2 when $N_\alpha=N_\beta-1.$

\noindent {\bf Theorem~2.4} {\it With the same notation as in
Theorem~2.1, assume that}
\item{(i)} $0<\beta<\alpha=\pi.$
\item{(ii)} $N_\alpha=N_{\beta}-1\ge 0.$
\item{(iii)} $\beta=\epsilon.$
\item{(iv)} $\sigma_{\text d}(H_{\alpha})=\sigma_{\text d}(\tilde H_{\gamma}).$
\item{(v)} {\it The intersection of $\sigma_{\text d}(H_\beta)$ and
$\sigma_{\text d}(\tilde H_\epsilon)$ contains at least $N_\alpha$ common
elements.}
\item{(vi)} $|F_\alpha(k)|=|\tilde F_{\gamma}(k)|$ {\it for} $k\in\bR.$

\noindent {\it Then, we have $\alpha=\gamma$ and $V=\tilde V.$
Equivalently said,
if $N_\alpha=N_{\beta}-1\ge 0$ and $0<\beta<\alpha=\pi$
then the data $\Cal D_4$ given in (2.7)
determines $V$ uniquely.}

We note that if $N_\alpha=0$ in Theorems~2.1-2.4 above, then $V$ itself is
reconstructed uniquely [cf. (5.1)-(5.11)] from
$|F_\alpha|$ without needing $\beta,$
$h_{\beta\alpha},$ or any possible eigenvalue of $H_\beta.$ The next result
is the analog of Theorem~2.1 but when $|F_\beta|$ is
known instead of $|F_\alpha|.$

\noindent {\bf Theorem~2.5} {\it With the same notation as in
Theorem~2.1, suppose that}
\item{(i)} $0<\beta<\alpha<\pi.$
\item{(ii)} $N_\alpha=N_{\beta}\ge 0.$
\item{(iii)} $h_{\beta\alpha}=h_{\epsilon\gamma}.$
\item{(iv)} $\sigma_{\text d}(H_{\alpha})=\sigma_{\text d}(\tilde H_{\gamma}).$
\item{(v)} $\sigma_{\text d}(H_\beta)=\sigma_{\text d}(\tilde H_\epsilon).$
\item{(vi)} $|F_\beta(k)|=|\tilde F_{\epsilon}(k)|$ {\it for} $k\in\bR.$

\noindent {\it Then, we have $\alpha=\gamma,$ $\beta=\epsilon,$
and $V=\tilde V.$ Equivalently,
if $N_\alpha=N_{\beta}\ge 0$ and $0<\beta<\alpha<\pi$
then the data $\Cal D_5$ given in (2.8)
uniquely determines
$\{V,\alpha,\beta\}.$}

We note that if $N_\beta=0$ in Theorem~2.5, then $V$ itself
is uniquely determined by $|F_\beta|$ without needing $h_{\beta\alpha}.$
The analog of Theorem~2.2 is given next when $|F_\beta|$ is
known instead of $|F_\alpha|;$ it is also the analog of
Theorem~2.5 when $\alpha=\pi.$

\noindent {\bf Theorem~2.6} {\it With the same notation as in
Theorem~2.1, assume that}
\item{(i)} $0<\beta<\alpha=\pi.$
\item{(ii)} $N_\alpha=N_{\beta}\ge 0.$
\item{(iii)} $\alpha=\gamma.$
\item{(iv)} $\sigma_{\text d}(H_{\alpha})=\sigma_{\text d}(\tilde H_{\gamma}).$
\item{(v)} $\sigma_{\text d}(H_\beta)=\sigma_{\text d}(\tilde H_\epsilon).$
\item{(vi)} $|F_\beta(k)|=|\tilde F_{\epsilon}(k)|$ {\it for} $k\in\bR.$

\noindent {\it Then, we have $\beta=\epsilon$ and $V=\tilde V.$
This is equivalent to saying that
if $N_\alpha=N_{\beta}\ge 0$ and $0<\beta<\alpha=\pi$
then the data $\Cal D_6$ defined in (2.9)
uniquely determines
$\{V,\beta\}.$}

In the next theorem we present
the analog of Theorem~2.3 but when $|F_\beta|$ is
known instead of $|F_\alpha|;$ equivalently, it is the analog of Theorem~2.5
when $N_\alpha=N_\beta-1.$

\noindent {\bf Theorem~2.7}
{\it With the same notation as in
Theorem~2.1, suppose that}
\item{(i)} $0<\beta<\alpha<\pi.$
\item{(ii)} $N_\alpha=N_{\beta}-1\ge 0.$
\item{(iii)} $\beta=\epsilon.$
\item{(iv)} $h_{\beta\alpha}=h_{\epsilon\gamma}.$
\item{(v)} $\sigma_{\text d}(H_{\alpha})=\sigma_{\text d}(\tilde H_{\gamma}).$
\item{(vi)} $\sigma_{\text d}(H_\beta)=\sigma_{\text d}(\tilde H_\epsilon).$
\item{(vii)} $|F_\beta(k)|=|\tilde F_{\epsilon}(k)|$ {\it for} $k\in\bR.$

\noindent {\it Then, we have $\alpha=\gamma$
and $V=\tilde V.$
Equivalently,
if $N_\alpha=N_{\beta}-1\ge 0$ and $0<\beta<\alpha<\pi$
then $\{V,\alpha\}$ is uniquely determined by
the data $\Cal D_7$ given in (2.10).}

Finally, we state the analog of Theorem~2.4 but when $|F_\beta|$ is
known instead of $|F_\alpha|;$ it is also
the analog of Theorem~2.7 when $\alpha=\pi.$

\noindent {\bf Theorem~2.8}
{\it With the same notation as in
Theorem~2.1, suppose that}
\item{(i)} $0<\beta<\alpha=\pi.$
\item{(ii)} $N_\alpha=N_{\beta}-1\ge 0.$
\item{(iii)} $\beta=\epsilon.$
\item{(iv)} $\alpha=\gamma.$
\item{(v)} $\sigma_{\text d}(H_{\alpha})=\sigma_{\text d}(\tilde H_{\gamma}).$
\item{(vi)} $\sigma_{\text d}(H_\beta)=\sigma_{\text d}(\tilde H_\epsilon).$
\item{(vii)} $|F_\beta(k)|=|\tilde F_{\epsilon}(k)|$ {\it for} $k\in\bR.$

\noindent {\it Then, we have $V=\tilde V.$
This is equivalent to saying that
if $N_\alpha=N_{\beta}-1\ge 0$ and $0<\beta<\alpha=\pi$
then $V$ is uniquely determined by
the data $\{\beta,|F_\beta(k)| \text{ for } k\in\bR,
\{\kappa_{\alpha j}\}_{j=1}^{N_\alpha},
\{\kappa_{\beta j}\}_{j=1}^{N_\beta}\}.$}

\vskip 10 pt
\noindent {\bf 3. PRELIMINARIES}
\vskip 3 pt

We first state some known results [7,8,10-12,19-22,28-31] that we need for the
proofs of Theorems~2.1-2.8.
Consider the Jost solution $f(k,x)$ to (1.2) with
the asymptotics in (1.6).
The properties of $f(k,x)$ are well understood.
For each fixed $x\in[0,+\infty),$ it is known that
$f(\cdot,x)$ and $f'(\cdot,x)$ are analytic in $\bCp$ and continuous in $\bCpb.$
Also, $f(k,0)$ and $f'(k,0)$ are real valued
if $k\in\bold I^+\cup\{0\},$
where $\bold I^+:=i(0,+\infty)$ is
the positive imaginary axis in $\bCp.$
Moreover, as
$k\to\infty$ in $\bCpb$ we have
$$f(k,0)=1-\ds\frac{1}{2ik}\int_0^\infty dx\,V(x)+o(1/k),\tag 3.1$$
$$f'(k,0)=ik-\ds\frac12\int_0^\infty dx\,V(x)+o(1),\tag 3.2$$
$$\ds\frac{f'(k,0)}{f(k,0)}=ik-\int_0^\infty dx\,
V(x)\,e^{2ikx}+o(1/k).\tag 3.3$$
Furthermore [32], we have as $k\to 0$ in $\bCpb,$
$$\ds\frac{f'(k,0)}{f(k,0)}=\ds\frac{f'(0,0)}{f(0,0)}+
\ds\frac{ik}{f(0,0)^2}+o(k),\qquad f(0,0)\ne 0,\tag 3.4$$
$$\ds\frac{f(k,0)}{f'(k,0)}=\ds\frac{f(0,0)}{f'(0,0)}-
\ds\frac{ik}{f'(0,0)^2}+o(k),\qquad f'(0,0)\ne 0.\tag 3.5$$
It is known that $f(\cdot,0)$ has a finite number of simple zeros in $\bCp,$
which correspond to the
eigenvalues of $H_\pi.$ The only
real zero of $f(\cdot,0)$ may occur as a simple zero at $k=0.$

The properties of the Jost function $F_\alpha$ defined in (1.5)
are also well understood [8]
and are summarized in the following proposition.

\noindent {\bf Proposition~3.1} {\it For $\alpha\in(0,\pi],$
let $F_\alpha$ be the Jost function
associated with a potential in the Faddeev class
and related to the boundary condition (1.3). Then, $F_\alpha$
is analytic in $\bCp$ and continuous in $\bCpb.$
Further, $F_\alpha$ has a finite number of zeros in
$\bCp$ and they are all located on
$\bold I^+.$
The zeros of $F_\alpha$ in $\bCpb$ are simple, and the only real zero of
$F_\alpha$ may occur as a simple zero at $k=0.$}

As stated below (1.6), we use $i\kappa_{\alpha j}$ to
denote the zeros of $F_\alpha$ in $\bCp,$ and we order them as
$0<\kappa_{\alpha 1}<\dots<\kappa_{\alpha N_\alpha}.$

\noindent {\bf Proposition~3.2} {\it Assume $V$ is in the Faddeev
class and $\alpha\in(0,\pi].$ Then, the corresponding
Jost function $F_\alpha$ can be
uniquely reconstructed from its amplitude given on $\bR$ and its
zeros in $\bCp.$ For $\alpha\in(0,\pi)$ we have
$$F_\alpha(k)=k\left(\ds\prod_{j=1}^{N_\alpha}\ds\frac{k-i\kappa_{\alpha
j}}{k+i\kappa_{\alpha j}}\right) \exp\left(\ds\frac{-1}{\pi
i}\int_{-\infty}^\infty dt\, \ds\frac{\log
(t/|F_\alpha(t)|)}{t-k-i0^+}\right), \qquad k\in\bCpb,\tag 3.6$$
and for $\alpha=\pi$ we get
$$F_\pi(k)=\left(\ds\prod_{j=1}^{N_\pi}
\ds\frac{k-i\kappa_{\pi j}}{k+i\kappa_{\pi j}}\right)
\exp\left(\ds\frac{1}{\pi i}\int_{-\infty}^\infty dt\,
\ds\frac{\log (|F_\pi(t)|)}{t-k-i0^+}\right),\qquad k\in\bCpb,\tag 3.7$$
where $i0^+$ indicates that the value for
$k\in\bR$ must be obtained as a limit from $\bCp.$}

\noindent PROOF: Let
$$G_\alpha(k):=\cases
\ds\frac{k}{F_\alpha(k)}\left(\ds\prod_{j=1}^{N_\alpha}
\ds\frac{k-i\kappa_{\alpha j}}{k+i\kappa_{\alpha j}}\right),
\qquad \alpha\in(0,\pi),\\
\stretch
F_\pi(k) \left(\ds\prod_{j=1}^{N_\pi}
\ds\frac{k+i\kappa_{\pi j}}{k-i\kappa_{\pi j}}\right),
\qquad \alpha=\pi.\endcases$$
With the help of (1.5) and Proposition~3.1, we see that
$G_\pi$ has no zeros in $\bCpb\setminus\{0\}$
and $\log G_\pi$ belongs to
the Hardy class of functions that are
analytic on $\bCp.$ From (3.1) we get
$$\log G_\pi(k)=O(1/k),\qquad k\to\infty {\text { in }} \bCpb.$$
Note that $f(0,0)$ and $f'(0,0)$ cannot simultaneously be zero
because this would imply $f(0,x)=0$ for
all $x\ge 0,$ contradicting (1.6).
Thus, when $f(0,0)=0,$
with the
help of (3.5) we get
$$\log G_\pi(k)=\log f'(k,0)+\log\left(
-ik/f'(0,0)^2\right)+O(1),\qquad k\to 0 {\text { in }}
\bCpb.$$
Consequently,
$$\int_{-\infty}^\infty dt\, |\log G_\pi(t+iz)|^2\le C,\qquad z\ge 0,$$
for some constant $C.$ Since $\log G_\pi$ is
analytic for $k\in\bCp$ and
$$\text{Re}[\log G_\pi(k)]=\log|G_\pi(k)|,\qquad k\in\bR,$$
it follows from the Schwarz integral formula
(see, e.g., Theorem~93 on p.~125 of [25]) that
$$\log G_\pi(k)=\ds\frac{1}{\pi i}\int_{-\infty}^\infty dt\,
\ds\frac{\log |G_\pi(t)|}{t-k},\qquad k\in\bCp.\tag 3.8$$
Moreover, $\log(G_\pi(t+iz))\to \log(G_\pi(t))$ as $z\to 0^+$ in the
$L^2$-sense and a.e. in $t.$ Consequently,
(3.7) follows from (3.8). We prove (3.6) in a similar way by
using the analyticity of $\log G_\alpha$
in $\bCp,$ (3.1)-(3.5), Proposition~3.1, and
$$\int_{-\infty}^\infty dt\, |\log G_\alpha(t+iz)|^2\le C,
\qquad z\ge 0,$$
for an appropriate constant $C.$ \qed

The large $k$-asymptotics of the Jost functions are treated
in the next proposition.

\noindent {\bf Proposition~3.3} {\it If $\alpha,\beta\in(0,\pi),$ then, as
$k\to \infty$ in $\bCpb,$ we have
$$F_\alpha(k)=k-i\left[\cot\alpha-\ds\frac12
\int_0^\infty dx\,V(x)\right]+o(1),\tag 3.9$$
$$F_\pi(k)=1-\ds\frac{1}{2ik}\int_0^\infty dx\,V(x)+o(1/k),\tag 3.10$$
$$F_\alpha(k)-F_\beta(k)=i\,h_{\beta\alpha}-\ds
\frac{h_{\beta\alpha}}{2k}\int_0^\infty dx\,V(x)+o(1/k),\tag 3.11$$
$$\ds\frac{F_\beta(k)}{F_\pi(k)}=k-i\,\cot\beta+
i\int_0^\infty dx\,
V(x)\,e^{2ikx}+o(1/k),\tag 3.12$$
$$\ds\frac{F_\pi(k)}{F_\beta(k)}=\ds\frac{1}{k}+
\ds\frac{i}{k^2}\left[\cot\beta-\int_0^\infty dx\,
V(x)\,e^{2ikx}\right]-\ds\frac{\cot^2\beta}{k^3}+o(1/k^3),\tag 3.13$$
$$\ds\frac{F_\alpha(k)}{F_\beta(k)}=1+\ds\frac{i\,h_{\beta \alpha}}{k}
-\ds\frac{h_{\beta\alpha}\,\cot\beta}{k^2}+o(1/k^2),\tag 3.14$$
where $h_{\beta \alpha}$ is the constant defined in (2.1).}

\noindent PROOF:  We obtain (3.9)-(3.14) directly
by using (3.1)-(3.3) in (1.5). \qed

Notice that $F_\alpha(k)$ is purely imaginary
for $k\in\bold I^+$ if $\alpha\in(0,\pi)$ and that
$F_\pi(k)$ is real for $k\in\bold I^+.$
Next, we analyze the small-$k$ asymptotics of the Jost function.
Since $f(0,0)$ and $f'(0,0)$ cannot be zero
at the same time,
with the help of (1.5) we see that if $F_\alpha(0)=0$ for any one value of
$\alpha\in(0,\pi)\setminus\{\pi/2\}$ then we must necessarily
have $f(0,0)\ne 0$ and $f'(0,0)\ne 0.$ Clearly, (1.5) also implies
that $F_\pi(0)=0$ if and only if $f(0,0)=0$ and $f'(0,0)\ne 0.$
Furthermore, from (1.5) we can conclude that
for $\alpha,\beta\in(0,\pi],$ if $\alpha\ne\beta$
then we cannot have $F_\alpha(0)=F_\beta(0)=0.$
Hence, in Propositions~3.4 and 3.5 below, we do not need to consider the trivial case
with $F_\alpha(0)=F_\beta(0)=0.$

\noindent {\bf Proposition~3.4} {\it Assume $\alpha,\beta\in(0,\pi).$
As $k\to 0$ in $\bCpb,$ we have}
$$\ds\frac{F_\alpha(k)}{F_\beta(k)}=\cases
\ds\frac{F_\alpha(0)}{F_\beta(0)}
-\ds\frac{ik\,h_{\beta \alpha}}{F_\beta(0)^2}+o(k),\qquad F_\beta(0)\ne 0,\\
\stretch
-\ds\frac{i}{k}\,\ds\frac{F_\alpha(0)^2}{h_{\beta\alpha}}[1+o(1)],
\qquad F_\beta(0)=0, \ \alpha\ne \beta.\endcases\tag 3.15$$

\noindent PROOF: Using
(1.5), (3.4), and (3.5), we expand $F_\alpha/F_\beta$ as $k\to 0$ in
$\bCpb$ and use (2.1) to simplify the result.
Note that if $F_\beta(0)=0,$ then with the help of (1.5) and (2.1) we
obtain $F_\alpha(0)=ih_{\beta\alpha} f(0,0),$ which
enables us to get the asymptotics in the second line of (3.15). \qed

\noindent {\bf Proposition~3.5} {\it Assume $\beta\in(0,\pi).$ As $k\to
0$ in $\bCpb,$ we have}
$$\ds\frac{F_\pi(k)}{F_\beta(k)}=\cases
\ds\frac{F_\pi(0)}{F_\beta(0)}
-\ds\frac{k}{F_\beta(0)^2}+o(k),\qquad F_\beta(0)\ne 0,\\
\stretch
\ds\frac{F_\pi(0)^2}{k}\left[1+o(1)\right],\qquad F_\beta(0)= 0.\endcases\tag 3.16$$
$$\ds\frac{F_\beta(k)}{F_\pi(k)}=\cases
\ds\frac{F_\beta(0)}{F_\pi(0)}
+\ds\frac{k}{F_\pi(0)^2}+o(k),\qquad F_\pi(0)\ne 0,\\
\stretch
-\ds\frac{F_\beta(0)^2}{k}[1+o(1)],\qquad F_\pi(0)=0.\endcases\tag 3.17$$

\noindent PROOF: Using (1.5), (3.4), and (3.5), we get
the expansion in the first line of
(3.16). Note that, if $F_\beta(0)=0,$ we must have $F_\pi(0)\ne 0$ and hence
we get the expansion in the second line of (3.16).
In a similar way, the first line of (3.17) is obtained from
(1.5) and (3.4), and the second line is obtained from (1.5) and (3.5) by noting that
$F_\beta(0)=-i\,f'(0,0)$ when  $F_\pi(0)=0.$ \qed

\noindent {\bf Proposition~3.6} {\it If $\alpha,\beta\in(0,\pi),$
then for $k\in\bR$ we have}
$$\text{Re}\left[\ds\frac{F_\pi(k)}{F_\beta(k)}\right]=\ds\frac{k}{|F_\beta(k)|^2},\quad
\text{Re}\left[\ds\frac{F_\beta(k)}{F_\pi(k)}\right]=\ds\frac{k}{|F_\pi(k)|^2},\quad
\text{Re}\left[\ds\frac{i\,F_\beta(k)}{F_\alpha(k)}\right]=
\ds\frac{k\,h_{\beta\alpha}}{|F_\alpha(k)|^2}.\tag 3.18$$

\noindent PROOF: The first two identities in (3.18) are obtained
directly from (1.5) and the well-known Wronskian identity [7,8,10,19-22,28-31]
$$f(k,0)\,\overline{f'(k,0)}-\overline{f(k,0)}\,f'(k,0)
=-2ik,\qquad k\in\bR,\tag 3.19$$
where an overbar denotes complex conjugation. To get the third
identity, we use (1.5), (2.1), and (3.19). \qed

\noindent {\bf Proposition~3.7} {\it Let $H_\alpha$ and $H_\beta$
be two realizations of the Schr\"odinger operator
for the potential $V$ in the Faddeev class
with respective boundary conditions $\alpha$ and $\beta,$ and
respective eigenvalues $\{-\kappa_{\alpha j}^2\}_{j=1}^{N_\alpha}$
and $\{-\kappa_{\beta j}^2\}_{j=1}^{N_{\beta}}.$ Assume that
$0<\beta<\alpha\le \pi.$ Then, $\sigma_{\text d}(H_\alpha)$ and
$\sigma_{\text d}(H_\beta)$ are disjoint, and either $N_{\beta}=N_\alpha$ or
$N_\beta=N_\alpha+1.$ In the former case we have
$$0<\kappa_{\alpha 1}<\kappa_{\beta 1}
<\kappa_{\alpha 2}<\kappa_{\beta 2}<\dots< \kappa_{\alpha
N_\alpha}<\kappa_{\beta N_\alpha},\tag 3.20$$
and in the latter
case we have
$$0<\kappa_{\beta 1}<\kappa_{\alpha 1}<
\kappa_{\beta 2}<\kappa_{\alpha 2}<\dots< \kappa_{\alpha
N_\alpha}<\kappa_{\beta (N_\alpha+1)}.\tag 3.21$$}

\noindent PROOF: First, let us prove that the eigenvalues of
$H_\alpha$ and $H_\beta$ cannot overlap. Recall that the
eigenvalues of $H_\alpha$ correspond to zeros of the Jost function
$F_\alpha$ in $\bCp.$ If $-\kappa^2$ were a common eigenvalue,
then we would have $F_\alpha(i\kappa)=F_\beta(i\kappa)=0.$ By (1.5), this
would imply $f(i\kappa,0)=f'(i\kappa,0)=0$ because
we assume $\alpha>\beta.$ This, however, would force
$f(i\kappa,x)=0$ for all $x\ge 0,$ which is incompatible with
(1.6). Next, let us prove that either $N_{\beta}=N_\alpha$ or
$N_\beta=N_\alpha+1,$ and that either (3.20) or (3.21) holds. The
quadratic form [33,34] $Q_\alpha$ associated with $H_\alpha$
is given by
$$Q_\alpha(\phi,\psi)=\langle\phi',\psi'\rangle+
\langle V\phi,\psi\rangle-\cot\alpha\cdot \phi(0)\cdot
\overline{\psi(0)},\qquad \alpha\in(0,\pi),$$
with domain
$W_{1,2}(0,+\infty),$ and
$$Q_\pi(\phi,\psi)=\langle\phi',\psi'\rangle+\langle V\phi,\psi\rangle,$$
with domain $W^{(0)}_{1,2}(0,+\infty).$ Here, we use
$\langle\cdot,\cdot\rangle$ for the standard scalar product in
$L^2(0,+\infty),$ $W_{1,2}(0,+\infty)$ for the standard Sobolev
space [35], and $W^{(0)}_{1,2}(0,+\infty)$ for that Sobolev space
with the Dirichlet boundary condition $\phi(0)=0.$ Note that
$W^{(0)}_{1,2}(0,+\infty)\subset W_{1,2}(0,+\infty).$ Since the
difference of the resolvents of $H_\alpha$ for different values of
$\alpha$ is a rank-one operator, it follows from the min-max
principle and the spectral mapping theorem [22] that the
eigenvalues of $H_\alpha$ and $H_\beta$ must interlace. Further,
we get $-\kappa_{\beta N_\beta}^2<-\kappa_{\alpha
N_\alpha}^2$ because $\beta<\alpha$ and $\sigma_{\text d}(H_\alpha)$ and
$\sigma_{\text d}(H_\beta)$ are disjoint. Thus, we must have either
$N_{\beta}=N_\alpha$ or $N_\beta=N_\alpha+1,$ and in the former
case (3.20) must hold and in the latter case (3.21) must hold. \qed

\noindent {\bf Proposition~3.8} {\it Assume $0<\beta<\alpha\le
\pi,$ and let $F_\alpha$ and $F_\beta$ be the Jost functions
associated with a potential in the Faddeev class with respective
boundary conditions $\alpha$ and $\beta.$ We have the following:}
\item{(i)} {\it If $F_\alpha(0)=0$ then
$N_\beta=N_\alpha+1.$}
\item{(ii)} {\it If $F_\beta(0)=0$ then $N_\beta=N_\alpha.$}

\noindent PROOF: From Propositions~3.1 and 3.7 we know that the zeros
of $F_\alpha$ and $F_\beta$ are simple and interlace on $\bold I^+$
and that either
$N_\beta=N_\alpha$ or $N_\beta=N_\alpha+1.$
The asymptotics of
$F_\alpha$ and $F_\beta$ as $k\to\infty$ on $\bold I^+$
are already known from Proposition~3.3;
by also analyzing the signs of
$F_\alpha$ and $F_\beta$ as $k\to 0$ on $\bold I^+,$
we can tell whether $N_\beta=N_\alpha$ or $N_\beta=N_\alpha+1.$
When $0<\beta<\alpha<\pi,$ we have
$N_\beta=N_\alpha$ if $F_\alpha/F_\beta$
remains positive (or approaches $0^+$ or $+\infty$) as
$k\to 0$ on $\bold I^+,$ and we have $N_\beta=N_\alpha+1$
if that sign remains negative (or approaches $0^-$ or $-\infty$).
When $0<\beta<\alpha=\pi,$ in the light of the
asymptotics in (3.13) as $k\to\infty$ on $\bold I^+,$
we have $N_\beta=N_\alpha$ if $iF_\pi/F_\beta$
remains positive (or approaches $0^+$ or $+\infty$) as
$k\to 0$ on $\bold I^+,$ and we have $N_\beta=N_\alpha+1$
if that sign remains negative (or approaches $0^-$ or $-\infty$).
In the former case of $\alpha\ne \pi,$ using the first line
of (3.15) with $F_\alpha(0)=0,$ we see that the sign of
$F_\alpha/F_\beta$ as $k\to 0$ on $\bold I^+$ coincides
with the sign of $h_{\beta\alpha}/F_\beta(0)^2,$ which is negative
due to the facts that $h_{\beta\alpha}>0$ and
$F_\beta(0)$ is purely imaginary. Thus,
(i) holds if $\alpha\in(0,\pi).$ On the other hand, if $\alpha=\pi,$ by
putting $F_\pi(0)=0$ in the first line of (3.16) and noting that
$F_\beta(0)$ is purely imaginary, we see that
the sign of $iF_\pi/F_\beta$
remains negative as $k\to 0$ on $\bold I^+.$ Thus, (i) is valid
also when $\alpha=\pi.$
Let us now turn to (ii). If $\alpha\in(0,\pi),$ by first interchanging
$\alpha$ and $\beta$ in the first line of (3.15) and
then by setting $F_\beta(0)=0$
there, we see that the sign of $F_\beta/F_\alpha$
as $k\to 0$ on $\bold I^+$ coincides
with the sign of $h_{\alpha\beta}/F_\alpha(0)^2,$
which is negative due to the facts
that $h_{\alpha\beta}=-h_{\beta\alpha}<0$
and $F_\alpha(0)$ is purely imaginary.
Thus, (ii) is proved when $\alpha\in(0,\pi).$
When $\alpha=\pi,$ from the
second line of (3.16) we see that
$iF_\pi/F_\beta$
remains positive as
$k\to 0$ on $\bold I^+,$ and
hence $N_\beta=N_\alpha$ if $F_\beta(0)=0,$
as stated in (ii). \qed

Next, we review certain known results [7,8,31,36-38] related to the
spectral function associated with $H_\alpha.$
Let $\varphi_\alpha(k,x)$ be the regular solution to
(1.2) satisfying the boundary conditions
$$\cases
\varphi_\alpha(k,0)=1,\quad \varphi'_\alpha(k,0)=-\cot\alpha,
\qquad \alpha\in(0,\pi),\\
\stretch
\varphi_\pi(k,0)=0,\quad \varphi'_\pi(k,0)=1.\endcases\tag 3.22$$
There is a monotone
increasing function $\rho_\alpha(\lambda)$ with $\lambda\in\bR,$
known as the spectral function, such that for any $g\in
L^2(0,+\infty),$
$$(U_\alpha g)(\lambda):=
\ds{\text{s-}}\ds\lim_{{n\to+\infty}}\ds\int_0^n dx\,
\varphi_\alpha(\sqrt{\lambda},x)\,g(x),$$
exists as a
strong limit in $L^2(\bR,d\rho_\alpha),$ and moreover the
following Parseval identity holds:
$$\langle g,h\rangle=\langle U_\alpha g,U_\alpha h\rangle,$$
where we recall that $\langle\cdot,\cdot\rangle$ is the standard
scalar product in $L^2(0,+\infty).$
The map $U_\alpha$ allows a
spectral representation of $H_\alpha.$ It follows from [7,8]
that
$$d\rho_\alpha(\lambda)=\cases
\ds\frac{\sqrt{\lambda}}{\pi}\,\ds\frac{1}
{|F_\alpha(\sqrt{\lambda})|^2}\,d\lambda,\qquad
\lambda >0,\\
\stretch
\ds\sum_{j=1}^{N_\alpha} g_{\alpha j}^2\,
\delta(\lambda+\kappa_{\alpha j}^2)\,d\lambda, \qquad \lambda <0,
\endcases\tag 3.23$$
where $\delta(\cdot)$ is the Dirac
delta distribution and the constants $g_{\alpha j}$ are given
(cf. [7,8]) by
$$g_{\alpha j}=:\cases \ds\frac{|f(i\kappa_{\alpha j},0)|}
{||f(i\kappa_{\alpha j},\cdot)||},\qquad \alpha\in(0,\pi),\\
\stretch
\ds\frac{|f'(i\kappa_{\pi j},0)|}{||f(i\kappa_{\pi
j},\cdot)||},\qquad \alpha=\pi,\endcases$$
with
$||\cdot||$ denoting the norm in $L^2(0,+\infty)$ and $f(k,x)$
being the Jost solution to (1.2). Note that the Marchenko norming
constants $m_{\alpha j}$ associated with the eigenvalues $-\kappa_{\alpha j}^2$
are defined as
$$m_{\alpha j}:=\ds\frac{1}{||f(i\kappa_{\alpha j},\cdot)||},\qquad
j=1,\dots,N_\alpha.$$
With the help of (4.2.19) of [8] and (1.5),
one can show that
$$||f(i\kappa_{\alpha j},\cdot)||^2=\cases
\ds\frac{1}{2\kappa_{\alpha j}}\,\dot F_\alpha(i\kappa_{\alpha j})\,
f(i\kappa_{\alpha j},0),\qquad \alpha\in(0,\pi),\\
\stretch
\ds\frac{i}{2\kappa_{\pi j}}\,\dot F_\pi(i\kappa_{\pi j})\,
f'(i\kappa_{\pi j},0),\qquad \alpha=\pi,\endcases\tag 3.24$$
with the overdot denoting the $k$-derivative.
Thus, if $\alpha\in(0,\pi),$ then
both $\{g_{\alpha j}\}_{j=1}^{N_\alpha}$ and $\{m_{\alpha j}\}_{j=1}^{N_\alpha}$
can be constructed once
$F_\alpha$ and $f(i\kappa_{\alpha j},0)$ are known.
On the other hand, if $\alpha=\pi,$ then we can construct
those norming constants when we know
$F_\pi$ and $f'(i\kappa_{\pi j},0).$
If $0<\beta<\alpha<\pi,$ as seen from (2.2),
once we know $F_\alpha,$ $F_\beta,$ and $h_{\beta\alpha},$
we can evaluate $f(k,0)$ and hence $f(i\kappa_{\alpha j},0);$
in particular, we get $F_\beta(i\kappa_{\alpha j})=-i\,
h_{\beta\alpha}\,f(i\kappa_{\alpha j},0).$
If $0<\beta<\alpha=\pi,$ from (1.5) it follows that
$f'(i\kappa_{\pi j},0)=i\,F_\beta(i\kappa_{\pi j}),$ and hence
knowledge of $F_\beta$ and $F_\pi$ allows us
to construct both the Gel'fand-Levitan and Marchenko norming constants.
We have
$$g_{\alpha j}=\cases
\ds\sqrt{\ds\frac{2i\kappa_{\alpha j}\,F_\beta(i\kappa_{\alpha j})}
{h_{\beta\alpha}\,\dot
F_\alpha(i\kappa_{\alpha j})}},\qquad 0<\beta<\alpha<\pi,\\
\stretch
\ds\sqrt{\ds\frac{2\,\kappa_{\pi j}\,F_\beta(i\kappa_{\pi j})}{\dot
F_\pi(i\kappa_{\pi j})}},\qquad 0<\beta<\alpha=\pi,\endcases\tag 3.25$$
$$m_{\alpha j}=\cases
\ds\sqrt{\ds\frac{-2i\kappa_{\alpha j}\,h_{\beta\alpha}}
{F_\beta(i\kappa_{\alpha j})\,\dot
F_\alpha(i\kappa_{\alpha j})}},\qquad 0<\beta<\alpha<\pi,\\
\stretch
\ds\sqrt{\ds\frac{-2\,\kappa_{\pi j}}
{F_\beta(i\kappa_{\pi j})\,\dot
F_\pi(i\kappa_{\pi j})}},\qquad 0<\beta<\alpha=\pi.\endcases\tag 3.26$$

\vskip 10 pt
\noindent {\bf 4. PROOFS OF THE MAIN THEOREMS}
\vskip 3 pt

In this section, we
present the proofs of Theorems~2.1-2.8.
In each proof, we describe how the boundary conditions
are uniquely reconstructed and how enough information
can be assembled for the
unique recovery of
the potential via the methods of Section~5.

\noindent PROOF OF THEOREM 2.1:
In this case we have $N_\beta=N_\alpha$ and
$0<\beta<\alpha<\pi.$
Since $|\tilde F_\gamma(k)|=|F_\alpha(k)|$ for $k\in\bR,$
it follows from (3.9) and (3.10) that
we have $\gamma<\pi;$ moreover, we get $\epsilon<\gamma$ because
$h_{\epsilon\gamma}=h_{\beta\alpha}>0.$
We would like to show
that our data $\Cal D_1$ given in (2.4)
uniquely reconstructs $V,$ $\alpha,$ and $\beta.$
Note that by Proposition~3.8(i),
we must have $F_\alpha(0)\ne 0.$ Define
$$\Lambda_1(k):=-i+i\,\ds\frac{F_\beta(k)}{F_\alpha(k)}
\ds\prod_{j=1}^{N_\alpha}\ds\frac{k^2+\kappa_{\alpha j}^2}{k^2+\kappa_{\beta j}^2}.\tag 4.1$$
 From the third formula in (3.18) it follows that
$$\text{Re}[\Lambda_1(k)]=\ds\frac{k\,h_{\beta\alpha}}{|F_\alpha(k)|^2}
\ds\prod_{j=1}^{N_\alpha}\ds\frac{k^2+\kappa_{\alpha j}^2}{k^2+\kappa_{\beta j}^2},
\qquad k\in\bR.\tag 4.2$$
The properties of
$F_\alpha$ and $F_\beta$ stated in Proposition~3.1
indicate that $\Lambda_1$ is analytic in $\bCp$ and continuous
in $\bCpb\setminus\{0\}.$ Using (3.14) with $\alpha$ and $\beta$
interchanged, from (4.1) we get
$$\Lambda_1(k)=\ds\frac{h_{\beta\alpha}}{k}+\ds\frac{i}{k^2}\left[
h_{\beta\alpha}\,\cot\alpha+\ds\sum_{j=1}^{N_\alpha}
(\kappa_{\alpha j}^2-
\kappa_{\beta j}^2)\right]
+o(1/k^2),\qquad k\to\infty \text{ in }\bCpb.\tag 4.3$$
As $k\to 0$ in $\bCpb,$ noting that
$F_\alpha(0)\ne 0$ and using
the first line in (3.15) with $\alpha$ and $\beta$ switched,
 from (4.1) we see that $\Lambda_1(k)=O(1)$ and hence
$\Lambda_1$ is continuous at $k=0.$
In terms of the data $\Cal D_1,$
we construct the right hand side of (4.2) and use it as input
to the Schwarz formula
$$\Lambda_1(k)=\ds\frac{1}{\pi i}\int_{-\infty}^\infty
\ds\frac{dt}{t-k-i0^+}\,
\text{Re}[\Lambda_1(t)],\qquad k\in\bCpb.\tag 4.4$$
Thus, $\Lambda_1$ is uniquely constructed.
Note that using $\Cal D_1$ and (4.3), we can recover
$\cot\alpha$ and hence $\alpha$ as well.
Then, $\cot\beta$ and hence $\beta$ can be recovered by using (2.1).
Our data also allows the construction of $F_\alpha$ in
$\bCpb$ via (3.6).
Then, having $F_\alpha$ and $\Lambda_1$ in hand, we obtain
$F_\beta$ from (4.1) as
$$F_\beta(k)=F_\alpha(k)\left[1-i\,\Lambda_1(k)\right]\ds\prod_{j=1}^{N_\alpha}
\ds\frac{k^2+\kappa_{\beta j}^2}{k^2+\kappa_{\alpha j}^2}.$$
Having $F_\alpha,$ $F_\beta,$ and $h_{\beta\alpha},$
we can reconstruct $V$ uniquely by using any one of the methods
described in Section~5.
Analogous to (2.4), let us define the data set $\tilde{\Cal D}_1$ as
$$\tilde{\Cal D}_1:=\{h_{\epsilon\gamma},|\tilde F_\gamma(k)| \text{ for } k\in\bR,
\{\tilde\kappa_{\gamma j}\}_{j=1}^{\tilde N_\gamma},
\{\tilde\kappa_{\epsilon j}\}_{j=1}^{\tilde N_\epsilon}\}.$$
Then, the uniqueness for $\Cal D_1\mapsto\{V,\alpha,\beta\}$
follows from the fact that $\tilde{\Cal D}_1=
\Cal D_1.$ \qed

\noindent PROOF OF THEOREM 2.2:  We have
$0<\beta<\alpha=\pi$ and $N_\pi=N_\beta.$
As in the proof of Theorem~2.1, we prove that
$\epsilon<\gamma=\pi.$ We cannot have
$F_\pi(0)=0$ as implied by Proposition~3.8(i).
We would like to show
that our data $\Cal D_2$ given in (2.5) uniquely reconstructs $V.$
Proceeding as in the proof of Theorem~2.1, let us define
$$\Lambda_2(k):=-1-\ds\frac{1}{k}\ds\frac{F_\beta(0)}{F_\pi(0)}
\ds\prod_{j=1}^{N_\pi}\ds\frac{\kappa_{\pi j}^2}{\kappa_{\beta j}^2}
+\ds\frac{1}{k}\ds\frac{F_\beta(k)}{F_\pi(k)}
\ds\prod_{j=1}^{N_\pi}\ds\frac{k^2+\kappa_{\pi j}^2}
{k^2+\kappa_{\beta j}^2}.\tag 4.5$$
Using the second identity of (3.18) in (4.5) and
noting that $F_\beta(0)$ is purely imaginary and
$F_\pi(0)$ is real, we see that
$$\text{Re}[\Lambda_2(k)]=-1+\ds\frac{1}{|F_\pi(k)|^2}
\ds\prod_{j=1}^{N_\pi}\ds\frac{k^2+\kappa_{\pi j}^2}
{k^2+\kappa_{\beta j}^2},
\qquad k\in\bR.\tag 4.6$$
Proposition~3.1
implies that $\Lambda_2$ is analytic in $\bCp$ and continuous
in $\bCpb\setminus\{0\}.$ Using (3.12) in (4.5) we get
$$\Lambda_2(k)=-\ds\frac{1}{k}\left[
i\,\cot\beta+\ds\frac{F_\beta(0)}{F_\pi(0)}
\ds\prod_{j=1}^{N_\pi}\ds\frac{\kappa_{\pi j}^2}
{\kappa_{\beta j}^2}\right]+
o(1/k),\qquad k\to\infty \text{ in }\bCpb.\tag 4.7$$
As $k\to 0$ in $\bCpb,$ since $F_\pi(0)\ne 0,$
with the help of the first line in
(3.17), from (4.5) we see that $\Lambda_2(k)=O(1)$
and hence $\Lambda_2$ is continuous at $k=0.$
Our data $\Cal D_2$ allows us to
construct $\Lambda_2$ by using the right hand side of (4.6) as input
to the appropriate Schwarz formula similar to (4.4).
Having constructed $\Lambda_2,$ using (4.7) we obtain
$$\ds\frac{F_\beta(0)}{F_\pi(0)}
\ds\prod_{j=1}^{N_\pi}\ds\frac{\kappa_{\pi j}^2}{\kappa_{\beta j}^2}=
-i\,\cot\beta-\ds\lim_{k\to\infty}[k\,\Lambda_2(k)],\tag 4.8$$
where the limit can be evaluated in any way in $\bCpb.$
Next, using (3.7) we construct $F_\pi.$
Then, using (4.5) and (4.8) we get
$$F_\beta(k)=k\,F_\pi(k)\left[\Lambda_2(k)+1-\ds\frac{i\,\cot\beta}{k}
-\ds\frac{1}{k}
\left(\ds\lim_{k\to\infty}[k\,\Lambda_2(k)]\right)\right]
\ds\prod_{j=1}^{N_\pi}\ds\frac{k^2+\kappa_{\beta j}^2}{k^2+\kappa_{\pi j}^2}.$$
Finally, having both $F_\pi$ and $F_\beta$ in hand,
$V$ can be reconstructed uniquely as indicated in Section~5. \qed

\noindent PROOF OF THEOREM 2.3: In this case we have
$N_\beta=N_\alpha+1$ and $0<\beta<\alpha<\pi.$ Arguing as in the
proof of Theorem~2.1, we get $\epsilon<\gamma<\pi.$
We would like to
show that our data $\Cal D_3$ defined in (2.6)
uniquely reconstructs $V,$ $\alpha,$
and $\beta.$ Notice that exactly
one of the $\kappa_{\beta j}$ is missing from our data.
Without loss of any generality, we can assume that
the missing element in $\Cal D_3$ is $\kappa_{\beta N_\beta}$ and use
$$\Cal D_3=\{h_{\beta\alpha},|F_\alpha(k)| \text{ for } k\in\bR,
\{\kappa_{\alpha j}\}_{j=1}^{N_\alpha},
\{\kappa_{\beta j}\}_{j=1}^{N_\alpha}\}.\tag 4.9$$
Our data allows us to construct $F_\alpha$ via (3.6).
By Proposition~3.8(ii), we see that $F_\beta(0)\ne 0.$ Define
$$\Lambda_3(k):=ik\,\ds\frac{F_\beta(k)}{F_\alpha(k)}
\ds\frac{\ds\prod_{j=1}^{N_\alpha}(k^2+\kappa_{\alpha j}^2)}
{\ds\prod_{j=1}^{N_\beta}(k^2+\kappa_{\beta j}^2)}.\tag 4.10$$
Proposition~3.1 indicates that $\Lambda_3$ is analytic
in $\bCp$ and continuous in $\bCpb.$ Using (3.14) with $\alpha$ and $\beta$
switched, we obtain
$$\Lambda_3(k)=\ds\frac{i}{k}+\ds\frac{h_{\beta\alpha}}{k^2}
+\ds\frac{i}{k^3}\left[h_{\beta\alpha}\cot\alpha
+\ds\sum_{j=1}^{N_\alpha}\kappa_{\alpha j}^2-
\ds\sum_{j=1}^{N_\beta}\kappa_{\beta j}^2\right]+o(1/k^3)
,\qquad k\to\infty {\text { in }} \bCpb.\tag 4.11$$
 From the third formula in (3.18) we get
$$\text{Re}[\Lambda_3(k)]=\ds\frac{h_{\beta\alpha}\,k^2}{|F_\alpha(k)|^2}
\ds\frac{\ds\prod_{j=1}^{N_\alpha}(k^2+\kappa_{\alpha j}^2)}
{\ds\prod_{j=1}^{N_\beta}(k^2+\kappa_{\beta j}^2)},
\qquad k\in\bR.\tag 4.12$$
If we had $\kappa_{\beta N_\beta}$ in $\Cal D_3,$ we would be able to
construct $\Lambda_3$ by using the right hand side of (4.12)
as input
into the appropriate Schwarz formula similar to (4.4)
and obtain
$$\Lambda_3(k)=\ds\frac{1}{\pi i}
\int_{-\infty}^\infty \ds\frac{dt}{t-k-i0^+}\,
\ds\frac{t^2}{t^2+\kappa_{\beta N_\beta}^2}
\ds\frac{h_{\beta\alpha}}{|F_\alpha(t)|^2}
\ds\prod_{j=1}^{N_\alpha}\ds\frac{t^2+\kappa_{\alpha j}^2}
{t^2+\kappa_{\beta j}^2},\qquad k\in\bCpb.\tag 4.13$$
However, since
$\kappa_{\beta N_\beta}$ is missing in our data, we proceed
in a slightly different manner. By replacing $\kappa_{\beta N_\beta}$
with an arbitrary positive parameter $\kappa$ on the right hand side of
(4.13), we obtain a one-parameter family of
functions
$$\Cal H(k,\kappa):=\ds\frac{1}{\pi i}
\int_{-\infty}^\infty \ds\frac{dt}{t-k-i0^+}\,
\ds\frac{t^2}{t^2+\kappa^2}
\ds\frac{h_{\beta\alpha}}{|F_\alpha(t)|^2}
\ds\prod_{j=1}^{N_\alpha}\ds\frac{t^2+\kappa_{\alpha j}^2}
{t^2+\kappa_{\beta j}^2},\qquad k\in\bCpb,\tag 4.14$$
that are analytic for $k\in\bCp$ and continuous for $k\in\bCpb.$
Note that
$\Cal H(k,\kappa_{\beta N_\beta})=\Lambda_3(k).$ Having
constructed $\Cal H(k,\kappa)$ containing $\kappa$ as a parameter,
we impose the restriction
$$\ds\lim_{k\to \infty} [k\,\Cal H(k,\kappa)]=i,\tag 4.15$$
so that, as seen from (4.11), the leading terms in the
large-$k$ asymptotics in $\Cal H(\cdot,\kappa)$ and
$\Lambda_3$ agree.
Provided we interpret
the limit as a nontangential limit in $\bCpb,$
we show in Proposition~4.1 that (4.15) has the unique positive solution
$\kappa=\kappa_{\beta N_\beta}.$ Having constructed
$\Cal H(k,\kappa)$ and $\kappa_{\beta N_\beta},$
we obtain $\Lambda_3(k)$ as
$\Cal H(k,\kappa_{\beta N_\beta}).$
Then, we construct $F_\beta$ via (4.10) as
$$F_\beta(k)=\ds\frac{1}{ik}\,F_\alpha(k)\,\Cal H(k,\kappa_{\beta N_\beta})
\,\ds\frac{\ds\prod_{j=1}^{N_\beta}(k^2+\kappa_{\beta j}^2)}
{\ds\prod_{j=1}^{N_\alpha}(k^2+\kappa_{\alpha j}^2)}.$$
Note that the value of $\cot\alpha$ can now be obtained from (4.11),
and then $\cot\beta$ can be computed via (2.1).
Thus, our data allows us to construct
$\alpha$ and $\beta.$ Having
$F_\alpha,$ $F_\beta,$ and $h_{\beta\alpha}$ in hand,
$V$ can be reconstructed uniquely via a method given in Section~5.
Alternatively, after obtaining $V,$ we can evaluate
$\alpha$ and $\beta$ with the help (3.9) and then (2.1). \qed

\noindent PROOF OF THEOREM 2.4: We have
$N_\beta=N_\alpha+1$ and $0<\beta<\alpha=\pi.$
As in the proof of Theorem~2.1 we prove that
$\epsilon<\gamma=\pi.$ We will
show that $\Cal D_4$ given in (2.7) uniquely reconstructs $V.$
As in the proof of Theorem~2.3,
without loss of any generality we can assume that
the missing element in $\Cal D_4$ is $\kappa_{\beta N_\beta}$ and use
$$\Cal D_4=\{\beta,|F_\pi(k)| \text{ for } k\in\bR,
\{\kappa_{\pi j}\}_{j=1}^{N_\pi},
\{\kappa_{\beta j}\}_{j=1}^{N_\pi}\}.\tag 4.16$$
We construct $F_\pi$ via (3.7).
 From Proposition~3.8(ii), we conclude that $F_\beta(0)\ne 0.$ Letting
$$\Lambda_4(k):=-1+k\,\ds\frac{F_\beta(k)}{F_\pi(k)}
\ds\frac{\ds\prod_{j=1}^{N_\pi}(k^2+\kappa_{\pi j}^2)}
{\ds\prod_{j=1}^{N_\beta}(k^2+\kappa_{\beta j}^2)},\tag 4.17$$
by Proposition~3.1 we observe
that $\Lambda_4$ is analytic
in $\bCp$ and continuous in $\bCpb.$ From (3.12) we obtain
$$\Lambda_4(k)=-\ds\frac{i\,\cot\beta}{k}+o(1/k)
,\qquad k\to\infty {\text { in }} \bCpb,\tag 4.18$$
and from the second identity in (3.18) we get
$$\text{Re}[\Lambda_4(k)]=-1+\ds\frac{k^2}{|F_\pi(k)|^2}
\ds\frac{\ds\prod_{j=1}^{N_\pi}(k^2+\kappa_{\pi j}^2)}
{\ds\prod_{j=1}^{N_\beta}(k^2+\kappa_{\beta j}^2)},
\qquad k\in\bR.\tag 4.19$$
If we had $\kappa_{\beta N_\beta}$ in $\Cal D_4,$ we could
construct $\Lambda_4$ by using (4.19)
as input
into the analog of (4.4)
and obtain
$$\Lambda_4(k)=\ds\frac{1}{\pi i}\int_{-\infty}^\infty
\ds\frac{dt}{t-k-i0^+}\,
\left[\ds\frac{t^2}{t^2+\kappa_{\beta N_\beta}^2}
\ds\frac{1}{|F_\pi(t)|^2}
\ds\prod_{j=1}^{N_\pi}\ds\frac{t^2+\kappa_{\pi j}^2}
{t^2+\kappa_{\beta j}^2}-1\right],\qquad k\in\bCpb.\tag 4.20$$
Since
$\kappa_{\beta N_\beta}$ is missing in our data, we proceed
as in the proof of Theorem~2.3. By replacing $\kappa_{\beta N_\beta}$
with an arbitrary positive parameter $\kappa$ on the right hand side of
(4.20), we obtain a one-parameter family of functions
$$\Cal H_\pi(k,\kappa):=
\ds\frac{1}{\pi i}\int_{-\infty}^\infty \ds\frac{dt}{t-k-i0^+}\,
\left[\ds\frac{t^2}{t^2+\kappa^2}
\ds\frac{1}{|F_\pi(t)|^2}
\ds\prod_{j=1}^{N_\pi}\ds\frac{t^2+\kappa_{\pi j}^2}
{t^2+\kappa_{\beta j}^2}-1\right],\qquad k\in\bCpb,$$
that are analytic for $k\in\bCp$ and continuous for $k\in\bCpb.$
Note that
$\Cal H_\pi(k,\kappa_{\beta N_\beta})=\Lambda_4(k).$ Having
constructed $\Cal H_\pi(\cdot,\kappa)$ containing $\kappa$ as a parameter,
we impose the restriction
$$\ds\lim_{k\to \infty} [k\,\Cal H_\pi(k,\kappa)]=-i\,\cot\beta,\tag 4.21$$
so that, as seen from (4.18), the leading terms in the
large-$k$ asymptotics in $\Cal H_\pi(\cdot,\kappa)$ and
$\Lambda_4$ agree.
We prove in Proposition~4.1 that (4.21) has the unique
positive solution $\kappa=\kappa_{\beta N_\beta}$ provided the limit
in (4.21) is a nontangential limit in $\bCpb.$ Having
$\Cal H_\pi(\cdot,\kappa)$ and $\kappa_{\beta N_\beta}$ in hand,
we obtain $\Lambda_4(k)$ as
$\Cal H_\pi(k,\kappa_{\beta N_\beta}).$
Then, $F_\beta$ is obtained via (4.17) as
$$F_\beta(k)=\ds\frac{1}{k}\,F_\pi(k)\,[\Cal H_\pi(k,\kappa_{\beta N_\beta})+1]
\,\ds\frac{\ds\prod_{j=1}^{N_\beta}(k^2+\kappa_{\beta j}^2)}
{\ds\prod_{j=1}^{N_\pi}(k^2+\kappa_{\pi j}^2)}.$$
Having found
$F_\pi$ and $F_\beta,$
$V$ can be reconstructed uniquely as explained in Section~5. \qed

\noindent{\bf Proposition~4.1} {Assume that each of
the data sets $\Cal D_3$ and $\Cal D_4$ given in
(4.9) and (4.16), respectively, is associated
with a potential in the Faddeev class. If the limits
in (4.15) and in (4.21) are interpreted
as nontangential limits in $\bCpb,$ then
(4.15) and (4.21) each have a unique positive solution, and that solution
is given by $\kappa=\kappa_{\beta N_\beta}.$}

\noindent PROOF: For the part of the proof related
to (4.15), we proceed as follows.
Define
$$I_1(k):=\int_{-\infty}^\infty dt\,
\ds\frac{k}{t-k-i0^+}\ds
\frac{\text{Re}[\Lambda_3(t)]}
{t^2+\kappa^2},\quad
I_2(k):=\int_{-\infty}^\infty dt\,
\ds\frac{t}{t-k-i0^+}\ds
\frac{\text{Re}[\Lambda_3(t)]}
{t^2+\kappa^2}.$$
With the help of (4.11) and (4.15) we see that the latter
is equivalent to
$$\ds\lim_{k\to \infty} [k\,\Cal H(k,\kappa)-k\,\Lambda_3(k)]=0,$$
and that (4.12)-(4.14) imply
$$k\,\Cal H(k,\kappa)-k\,\Lambda_3(k)=
\ds\frac{\kappa_{\beta N_\beta}^2-\kappa^2}{\pi i}\,I_1(k),$$
and hence our proof will be completed by showing that
the nontangential limit of $I_1(k)$ exists and is nonzero.
We note that
$$I_1(k)-I_2(k)=-\int_{-\infty}^\infty dt\,
\ds\frac{\text{Re}[\Lambda_3(t)]}{t^2+\kappa^2}.\tag 4.22$$
Writing $k$ in terms of its real and imaginary parts as
$k:=k_R+ik_I,$ from (3.9) and (4.12) we obtain
$$|I_2(k)|\le C \int_{-\infty}^\infty dt\,
\ds\frac{|t|}{\sqrt{(t-k_R)^2+k_I^2}}\,\ds\frac{1}
{(t^2+\kappa^2)(t^2+\kappa_{\beta N_\beta}^2)},$$
for an appropriate constant $C.$ With the help of the estimate
$$\ds\frac{1}{\sqrt{(t-k_R)^2+k_I^2}}\le\cases
\ds\frac{1}{k_I},\qquad |t|\ge |k_R|/2,\\
\ds\frac{1}{\sqrt{k_R^2/4+k_I^2}},\qquad
|t|\le |k_R|/2,\endcases$$
we get $I_2(k)=o(1)$ as $k\to\infty$ in
$\bCpb$ provided $k_I\ge \delta_1$
for some positive $\delta_1.$
Using the facts [cf. (4.12)] that
$\text{Re}[\Lambda_3(t)]$ is bounded
on $\bR$ and is positive when $t\ne 0,$
we conclude from (4.22) that
the nontangential limit $\lim_{k\to \infty}I_1(k)$
exists and is negative.

Arguing as above, we prove that (4.21) has the unique
positive solution $\kappa=\kappa_{\beta N_\beta}$ provided that
the nontangential limit $\lim_{k \to \infty} I(k)$
in $\bCpb$ is zero, where we have defined
$$\ds I(k):=\int_{-\infty}^\infty dt\,
\ds\frac{t}{t-k-i0^+}\ds \frac{\text{Re}[\Lambda_4(t)+1]}
{t^2+\kappa^2}.$$
For any $\Upsilon>0,$  let us write
$I(k)=I_3(k)+I_4(k)$ with
$$I_3(k):=\int_{|t| \ge \Upsilon} dt\, \ds\frac{t}{t-k-i0^+}\ds
\frac{\text{Re}[\Lambda_4(t)+1]} {t^2+\kappa^2},\quad
I_4(k):=\int_{|t|\le \Upsilon} dt\, \ds\frac{t}{t-k-i0^+}\ds
\frac{\text{Re}[\Lambda_4(t)+1]} {t^2+\kappa^2}.$$
By the Schwarz inequality, we have
$$|I_3(k)|^2 \le C \left(\int_{-\infty}^\infty
\frac{dt}{t^2+k_I^2}\right)\left(\int_{|t|\ge \Upsilon}dt\,
\frac{t^2}{(t^2+\kappa^2)^2}\right),$$
where $C$ is an appropriate constant [cf. (4.19)].
Thus, given $\delta_2,\delta_3>0$
we can take $\Upsilon$ large enough so that
$|I_3(k)|\le \delta_2$
for all $k\in\bCp$ with $k_I\ge \delta_3.$ Moreover, with $\Upsilon$ fixed
as above, for $|k_R|>2\Upsilon$ we get $|I_4(k)|\le  C\Upsilon/(|k_R|+k_I)$
for an appropriate constant $C.$
Hence the nontangential limit $\lim_{k \to \infty} I(k)$ is zero. \qed

\noindent PROOF OF THEOREM 2.5: In this case we have
$N_\beta=N_\alpha$ and $0<\beta<\alpha<\pi.$
As in the proof of Theorem~2.1, we prove that
$\epsilon<\gamma<\pi.$ We would like to show
that $\Cal D_5$ given in (2.8)
uniquely reconstructs $V,$ $\alpha,$ and
$\beta.$
Let
$$\Lambda_5(k):=ik-h_{\beta\alpha}-
ik\,\ds\frac{F_\alpha(k)}{F_\beta(k)}
\ds\prod_{j=1}^{N_\beta}\ds\frac{k^2+\kappa_{\beta
j}^2}{k^2+\kappa_{\alpha j}^2}.\tag 4.23$$
Using the third
identity in (3.18) with $\alpha$ and $\beta$ switched, from (4.23)
it follows that
$$\text{Re}[\Lambda_5(k)]=-h_{\beta\alpha}+\ds\frac{k^2\,h_{\beta\alpha}}{|F_\beta(k)|^2}
\ds\prod_{j=1}^{N_\beta}\ds\frac{k^2+\kappa_{\beta
j}^2}{k^2+\kappa_{\alpha j}^2},
\qquad k\in\bR,\tag 4.24$$
where we have also used
$h_{\beta\alpha}=-h_{\alpha\beta}.$
The properties of
$F_\alpha$ and $F_\beta$ stated in Proposition~3.1 allow us to
conclude that $\Lambda_5$ is analytic in $\bCp$ and continuous
in $\bCpb\setminus\{0\}.$ With the help of (3.14), from (4.23) we
get
$$\Lambda_5(k)=\ds\frac{i}{k}\left[h_{\beta\alpha}
\cot\beta+\ds\sum_{j=1}^{N_\beta}
(\kappa_{\alpha j}^2-\kappa_{\beta j}^2)\right]+ o(1/k),\qquad
k\to\infty \text{ in } \bCpb.\tag 4.25$$
As $k\to 0$ in $\bCpb,$
using (3.15) in (4.23) we see that $\Lambda_5(k)=O(1)$ regardless
of whether $F_\beta(0)=0$ or not, and hence $\Lambda_5$ is
continuous at $k=0.$ Then, in terms of $\Cal D_5,$ we construct
$\Lambda_5$ with the right hand side of (4.24) as input to the
appropriate Schwarz formula analogous to (4.4).
Using (4.25) we get the value of $\cot\beta$ and hence $\beta.$
Then, with the help of (2.1) we get the value of $\alpha.$
Next, using (3.6)
our data allows us to construct $F_\beta$ in $\bCpb.$ Then,
having $F_\beta$ and $\Lambda_5$ in hand, we obtain
$F_\alpha$ via (4.23) as
$$F_\alpha(k)=\ds\frac{i}{k}\,
F_\beta(k)\left[h_{\beta\alpha}-ik+
\Lambda_5(k)\right]\ds\prod_{j=1}^{N_\beta}
\ds\frac{k^2+\kappa_{\alpha j}^2}{k^2+\kappa_{\beta j}^2}.$$
Finally, having $F_\alpha,$ $F_\beta,$ and
$h_{\beta\alpha}$ in hand, $V$ is reconstructed
uniquely as indicated in Section~5. \qed

\noindent PROOF OF THEOREM 2.6: We are in the case
$0<\beta<\alpha=\pi$ and $N_\pi=N_\beta.$ As in the proof of
Theorem~2.1 we establish $\epsilon<\gamma=\pi,$
and we note that we cannot have $F_\pi(0)=0$ due to the assumption
$N_\pi=N_\beta.$ We will show that $\Cal D_6$
defined in (2.9) uniquely reconstructs $V$ and $\beta.$
Proceeding as in the proof of
Theorem~2.1, let us define
$$\Lambda_6(k):=-1+k\,
\ds\frac{F_\pi(k)}{F_\beta(k)}
\ds\prod_{j=1}^{N_\beta}\ds\frac{k^2+\kappa_{\beta
j}^2}{k^2+\kappa_{\pi j}^2}.\tag 4.26$$
Using the first identity
of (3.18) in (4.26), we obtain
$$\text{Re}[\Lambda_6(k)]=-1+\ds\frac{k^2}{|F_\beta(k)|^2}
\ds\prod_{j=1}^{N_\beta}\ds\frac{k^2+\kappa_{\beta
j}^2}{k^2+\kappa_{\pi j}^2},
\qquad k\in\bR.\tag 4.27$$
Proposition~3.1 implies
that $\Lambda_6$ is analytic in $\bCp$ and continuous in
$\bCpb\setminus\{0\}.$ Using (3.13) in (4.26) we get
$$\Lambda_6(k)=\ds\frac{i\,\cot\beta}{k}+
o(1/k),\qquad k\to\infty \text{ in }\bCpb.\tag 4.28$$
As $k\to 0$
in $\bCpb,$ using (3.16) in (4.26) we see that $\Lambda_6(k)=O(1)$
regardless of whether $F_\beta(0)=0$ or not, and hence
$\Lambda_6$ remains continuous at $k=0.$ Our data $\Cal D_6$
allows us to construct $\Lambda_6$ with the right hand side of
(4.27) used as input to the appropriate Schwarz formula, which is
the analog of (4.4). Having constructed $\Lambda_6,$ we recover
$\beta$ with the help of (4.28). Next, using (3.6)
we construct $F_\beta$ in $\bCpb,$ and from (4.26) we get
$$F_\pi(k)=\ds\frac{1}{k}\,F_\beta(k)\,[\Lambda_6(k)+1]\,
\ds\prod_{j=1}^{N_\beta}\ds\frac{k^2+\kappa_{\pi
j}^2}{k^2+\kappa_{\beta j}^2}.$$
Then, having both
$F_\pi$ and $F_\beta$ in hand, $V$ can be reconstructed
uniquely as shown in Section~5. \qed

\noindent PROOF OF THEOREM 2.7: This is the case
$N_\beta=N_\alpha+1$ and $0<\beta<\alpha<\pi.$ We prove
$\epsilon<\gamma<\pi$ as in the proof of Theorem~2.1. We would like to
show that our data $\Cal D_7$ given in (2.10) uniquely reconstructs $V$ and
$\alpha.$
Since $\Cal D_7$ contains $\beta$ and
$h_{\beta\alpha},$ we get $\alpha$ from (2.1). In this case Proposition~3.8(ii)
implies that $F_\beta(0)\ne 0.$ Define
$$\Lambda_7(k):=-ik+h_{\beta\alpha}-
\ds\frac{i}{k}\,\ds\frac{F_\alpha(0)}{F_\beta(0)}
\ds\frac{\ds\prod_{j=1}^{N_\beta}\kappa_{\beta j}^2}
{\ds\prod_{j=1}^{N_\beta-1}\kappa_{\alpha j}^2}+
\ds\frac{i}{k}\,\ds\frac{F_\alpha(k)}{F_\beta(k)}
\ds\frac{\ds\prod_{j=1}^{N_\beta}(k^2+\kappa_{\beta j}^2)}
{\ds\prod_{j=1}^{N_\beta-1}(k^2+\kappa_{\alpha j}^2)}.\tag 4.29$$
Using the third identity in (3.18) with $\alpha$ and $\beta$
switched, from (4.29) we get
$$\text{Re}[\Lambda_7(k)]=
h_{\beta\alpha}-\ds\frac{h_{\beta\alpha}}{|F_\beta(k)|^2}
\ds\frac{\ds\prod_{j=1}^{N_\beta}(k^2+\kappa_{\beta j}^2)}
{\ds\prod_{j=1}^{N_\beta-1}(k^2+\kappa_{\alpha j}^2)},
\qquad k\in\bR,\tag 4.30$$
where we have also used $h_{\beta\alpha}=-h_{\alpha\beta}.$
Proposition~3.1 indicates that $\Lambda_7$ is analytic in
$\bCp$ and continuous in $\bCpb\setminus\{0\}.$ With the help of
(3.14), as $k\to\infty$ in $\bCpb$ from (4.29) we get
$$\Lambda_7(k)=\ds\frac{i}{k}\left[-h_{\beta\alpha}\,\cot\beta+
\ds\sum_{j=1}^{N_\beta}\kappa_{\beta
j}^2-\ds\sum_{j=1}^{N_\beta-1}\kappa_{\alpha j}^2
-\ds\frac{F_\alpha(0)}{F_\beta(0)}
\ds\frac{\ds\prod_{j=1}^{N_\beta}\kappa_{\beta j}^2}
{\ds\prod_{j=1}^{N_\beta-1}\kappa_{\alpha j}^2}\right]+o(1/k).$$
Setting
$$P(k):=
ik\,\Lambda_7(k)-h_{\beta\alpha}\,\cot\beta+
\ds\sum_{j=1}^{N_\beta}\kappa_{\beta
j}^2-\ds\sum_{j=1}^{N_\beta-1}\kappa_{\alpha j}^2,\tag 4.31$$
we see that
$$\ds\frac{F_\alpha(0)}{F_\beta(0)}
\ds\frac{\ds\prod_{j=1}^{N_\beta}\kappa_{\beta j}^2}
{\ds\prod_{j=1}^{N_\beta-1}\kappa_{\alpha
j}^2}=\ds\lim_{k\to\infty} P(k),\tag 4.32$$
where the limit can be
obtained in any manner in $\bCpb.$ As
$k\to 0$ in $\bCpb,$ using the first line of (3.15) in (4.29) we
see that $\Lambda_7(k)=O(1)$ regardless of $F_\alpha(0)=0$ or $F_\alpha(0)\ne 0,$
and hence $\Lambda_7$ is continuous at $k=0.$ Then, the data $\Cal D_7$
allows us to construct $\Lambda_7$ with the right hand
side of (4.30) used as input to the appropriate Schwarz formula,
which is the analog of (4.4). Next, using (3.6) we construct
$F_\beta$ in $\bCpb.$ Consequently, using (4.32) in (4.29) we
are able to obtain $F_\alpha$ as
$$F_\alpha(k)=\ds\frac{k}{i}\,
F_\beta(k)\left[ik-h_{\beta\alpha}+\Lambda_7(k)+\ds\frac{i}{k}\,\left(
\ds\lim_{k\to\infty} P(k)\right)
\right]\ds\frac{\ds\prod_{j=1}^{N_\beta-1} (k^2+\kappa_{\alpha
j}^2)}{ \ds\prod_{j=1}^{N_\beta}(k^2+\kappa_{\beta j}^2)},$$
where $P$ is as given in (4.31). Finally, having
$F_\alpha,$ $F_\beta,$ and $h_{\beta\alpha}$ in hand, $V$
can be reconstructed uniquely as outlined in
Section~5. \qed

\noindent PROOF OF THEOREM 2.8: We have
$N_\beta=N_\alpha+1$ with $0<\beta<\alpha=\pi.$
 From (i), (iii), and (iv) we conclude
that $\epsilon<\gamma=\pi.$ We will
show that $\Cal D_8$ given in (2.11)
uniquely reconstructs $V.$
Define
$$\Lambda_8(k):=-1-
\ds\frac{1}{k}\,\ds\frac{F_\pi(0)}{F_\beta(0)}
\ds\frac{\ds\prod_{j=1}^{N_\beta}\kappa_{\beta j}^2}
{\ds\prod_{j=1}^{N_\beta-1}\kappa_{\alpha j}^2}+
\ds\frac{1}{k}\,\ds\frac{F_\pi(k)}{F_\beta(k)}
\ds\frac{\ds\prod_{j=1}^{N_\beta}(k^2+\kappa_{\beta j}^2)}
{\ds\prod_{j=1}^{N_\beta-1}(k^2+\kappa_{\alpha j}^2)}.\tag 4.33$$
Via (3.6) we construct $F_\beta$ in $\bCpb.$ Using the first
identity of (3.18) in (4.33) and noting that $F_\beta(0)$ is
purely imaginary and $F_\pi(0)$ is real, it follows that
$$\text{Re}[\Lambda_8(k)]=-1+\ds\frac{1}{|F_\beta(k)|^2}
\ds\frac{\ds\prod_{j=1}^{N_\beta}(k^2+\kappa_{\beta j}^2)}
{\ds\prod_{j=1}^{N_\beta-1}(k^2+\kappa_{\alpha j}^2)},
\qquad k\in\bR.\tag 4.34$$
Proposition~3.1 implies that $\Lambda_8$ is analytic in $\bCp$
and continuous in $\bCpb\setminus\{0\}.$ With the
help of (3.13), from (4.33) we get
$$\Lambda_8(k)=\ds\frac{1}{k}\left[i\,\cot\beta-\ds\frac{F_\pi(0)}{F_\beta(0)}
\ds\frac{\ds\prod_{j=1}^{N_\beta}\kappa_{\beta j}^2}
{\ds\prod_{j=1}^{N_\beta-1}\kappa_{\alpha j}^2}\right]+
o(1/k),\qquad k\to\infty {\text { in }} \bCpb.\tag 4.35$$
Again we
have $F_\beta(0)\ne 0$ because of Proposition~3.8(ii). As $k\to 0$
in $\bCpb,$ using the first line of (3.16) in (4.33) we see that
$\Lambda_8(k)=O(1)$ and hence $\Lambda_8$ is continuous at
$k=0.$ Now from the data $\Cal D_8,$ we construct $\Lambda_8$
with the right hand side of (4.34) used as input to the
appropriate Schwarz formula similar to (4.4). Then, with the help
of (4.33) and (4.35), we construct $F_\pi$ via
$$F_\pi(k)=k\,
F_\beta(k)\left[1+\Lambda_8(k)+\ds\frac{i\,\cot\beta}{k}-\ds\frac
{1}{k}\,\left(\ds\lim_{k\to\infty}
[k\,\Lambda_8(k)]\right)\right]\ds\frac{\ds\prod_{j=1}^{N_\beta-1}
(k^2+\kappa_{\alpha j}^2)}{
\ds\prod_{j=1}^{N_\beta}(k^2+\kappa_{\beta j}^2)}.$$
Finally, having both $F_\pi$ and $F_\beta$ in hand, $V$ can be
reconstructed uniquely as outlined in Section~5. \qed

\vskip 10 pt
\noindent {\bf 5. RECONSTRUCTION OF THE POTENTIAL}
\vskip 3 pt

In this section we outline several methods via which the potential can be
uniquely reconstructed from each of the data
sets $\Cal D_1,\dots,\Cal D_8$ given in (2.4)-(2.11).
These methods include the Gel'fand-Levitan method
[7,8,10,11,31] and the Marchenko
method [7,8,10,12,31] for the half-line inverse scattering problem,
the Faddeev-Marchenko [7,10,19-22] method
and several other methods [10,21] used
to solve the full-line inverse scattering problem.
We will show that each of $\Cal D_1,\dots,\Cal D_8$
constructs $\Cal G_\alpha,$ $\Cal M_\alpha,$ and
$\Cal F$ defined in (5.1), (5.7), and (5.14), respectively.
If we have $F_\alpha,$ $F_\beta,$ and $h_{\beta\alpha}$
in hand, the norming
constants $g_{\alpha j}$ and $m_{\alpha j}$ are constructed
via the first line of (3.25) and of (3.26), respectively. Thus,
each of $\Cal D_1,$
$\Cal D_3,$
$\Cal D_5,$
and $\Cal D_7$ yields $\Cal G_\alpha$ and $\Cal M_\alpha.$
On the other hand, if we have
$F_\pi$ and $F_\beta$
in hand, the norming
constants $g_{\pi j}$ and $m_{\pi j}$ are constructed
as in the second line of (3.25) and of (3.26), respectively. Thus,
each of $\Cal D_2,$
$\Cal D_4,$
$\Cal D_6,$
and $\Cal D_8$ yields $\Cal G_\pi$ and $\Cal M_\pi.$
The construction of $\Cal F$ from $\Cal D_1,\dots,\Cal D_8$
is achieved by using (5.18)-(5.23).

The data set $\Cal G_\alpha$
used as input to the Gel'fand-Levitan method is given by
$$\Cal G_\alpha:=\{|F_\alpha(k)|
\text{ for } k\in\bR,\{\kappa_{\alpha j}\}_{j=1}^{N_\alpha},
\{g_{\alpha j}\}_{j=1}^{N_\alpha}\},\qquad \alpha\in(0,\pi].\tag 5.1$$
It allows us to reconstruct the
corresponding regular
solution $\varphi_\alpha(k,x)$ uniquely as [cf. (3.22)]
$$\varphi_\alpha(k,x)=\cases
\cos kx+\ds\int_0^x dy\,
A_\alpha(x,y)\,\cos ky,\qquad \alpha\in(0,\pi),\\
\stretch
\ds\frac{\sin kx}{k}+\int_0^x dy\,
A_\pi(x,y)\,\ds\frac{\sin ky}{k},\qquad \alpha=\pi,\endcases\tag 5.2$$
and the
corresponding potential $V$ uniquely as
$$V(x)=2\,\ds\frac{d}{dx}A_\alpha(x,x^-),\qquad \alpha\in(0,\pi],\tag 5.3$$
where $A_\alpha(x,y)$ is obtained by solving
the Gel'fand-Levitan integral equation [7,8,10,11]
$$A_\alpha(x,y)+G_\alpha(x,y)+\int_0^x dz\,
G_\alpha(y,z)\,A_\alpha(x,z)=0,\qquad 0<y<x,\tag 5.4$$
with the kernel $G_\alpha(x,y)$ for $\alpha\in(0,\pi)$ given by
$$G_\alpha(x,y):=\ds\frac{1}{\pi}\int_{-\infty}^\infty
dk\,\left[\ds\frac{k^2}{|F_\alpha(k)|^2}-1\right]\left(\cos
kx\right) \left(\cos ky\right)+ \ds\sum_{j=1}^{N_{\alpha}}
g_{\alpha j}^2 \left(\cosh
\kappa_{\alpha j}x\right)\left(\cosh \kappa_{\alpha j}y\right),\tag 5.5$$
and the kernel $G_\pi(x,y)$ given by
$$G_\pi(x,y):=
\ds\frac{1}{\pi}\int_{-\infty}^\infty
dk\,\left[\ds\frac{1}{|F_\pi(k)|^2}-1\right]\left(\sin
kx\right) \left(\sin ky\right)+ \ds\sum_{j=1}^{N_{\pi}}
\ds\frac{g_{\pi j}^2}{\kappa_{\pi j}^2} \left(\sinh
\kappa_{\pi j}x\right)\left(\sinh \kappa_{\pi j}y\right).
\tag 5.6$$
We note that, with the help of
(3.9) and (3.10), it is possible to tell
whether we have $\alpha<\pi$ or
$\alpha=\pi.$ When $\alpha<\pi,$
we observe that $\alpha$ is readily obtained from the solution to (5.4)
because (3.22) and
(5.2) imply that $\cot\alpha=-A_\alpha(0,0).$

The data $\Cal M_\alpha$
used as input to the
Marchenko method is given by
$$\Cal M_\alpha:=\{S_\alpha(k) \text{ for } k\in\bR,\{\kappa_{\alpha j}\}_
{j=1}^{N_\alpha},\{m_{\alpha j}\}_{j=1}^{N_\alpha}\},
\qquad \alpha\in(0,\pi],\tag 5.7$$
where $S_\alpha$ is the scattering matrix defined in (1.7).
Given $\Cal M_\alpha,$ we are able to
reconstruct the
corresponding Jost
solution $f(k,x)$ uniquely as [cf. (1.6)]
$$f(k,x)=e^{ikx}+\int_x^\infty dy\,K(x,y)\,e^{iky},\tag 5.8$$
and the potential $V$ uniquely as
$$V(x)=-2\,\ds\frac{d}{dx}K(x,x^+),\tag 5.9$$
where $K(x,y)$ is obtained by solving
the Marchenko integral equation [7,8,10,12]
$$K(x,y)+M_\alpha(x+y)+\int_x^\infty dz\,M_\alpha(y+z)\,K(x,z)=0,\qquad
0<x<y,\tag 5.10$$
with the kernel
$$M_\alpha(y):=\cases\ds\frac{1}{2\pi}
\int_{-\infty}^\infty dk\,[S_\alpha(k)-1]\,e^{iky}
+\ds\sum_{j=1}^{N_{\alpha}}m_{\alpha j}^2\,e^{-\kappa_{\alpha j}y},
\qquad \alpha\in(0,\pi),\\
\stretch
\ds\frac{1}{2\pi}
\int_{-\infty}^\infty dk\,[1-S_\pi(k)]\,e^{iky}
+\ds\sum_{j=1}^{N_{\pi}}m_{\pi j}^2\,e^{-\kappa_{\pi j}y},
\qquad \alpha=\pi.\endcases\tag 5.11$$
Note that the
solution $K(x,y)$ to (5.10)
is the same for all $\alpha\in(0,\pi],$ whereas
the solution $A_\alpha(x,y)$ to (5.4) depends on $\alpha.$
This is not surprising because $K(x,y)$ is related [cf. (5.8)] to
the Fourier transform
of the Jost solution $f(k,x)$, which is independent of
$\alpha,$ whereas $A_\alpha(x,y)$ is related [cf. (5.2)] to the
Fourier transform of the regular solution $\varphi_\alpha(k,x),$
which depends on $\alpha.$ Let us also remark on the limiting values
$A_\alpha(x,x^-)$ and $K(x,x^+)$ appearing in (5.3) and
(5.9), respectively. If we invert the Fourier transforms
given in (5.2) and (5.8), we obtain
$A_\alpha(x,y)=0$ for $y>x$ and $K(x,y)=0$ for $y<x.$
To emphasize the jump discontinuities in these functions
when $y=x,$ we use the appropriate limiting values in (5.3) and
(5.9), even though those limits are not always
explicitly indicated in the literature (cf. [7,8,10]).

The potential $V$ can alternatively
be reconstructed by using the Gel'fand-Levitan method or the Marchenko
method in the Dirichlet case. This can be done as follows.
If we have $F_\alpha,$ $F_\beta,$ and $h_{\beta\alpha}$ for
some $\alpha,\beta\in(0,\pi)$ with $\alpha\ne \beta,$ then
by using (2.2) we can construct $F_\pi(k):=f(k,0).$
Having $F_\pi$ in hand, we also have
the $\kappa_{\pi j}$ for $j=1,\dots,N.$ Finally, the Gel'fand-Levitan norming
constants $g_{\pi j}$ and the Marchenko norming
constants $m_{\pi j}$ can be constructed by using the second line
of (3.25) and of (3.26), respectively.

One can also reconstruct $V$ by viewing it as the potential
in the full-line Schr\"odinger equation with $V\equiv 0$ for $x<0.$
Recall that the left Jost solution $f_{\text{l}}(k,x)$ and
the right Jost solution $f_{\text{r}}(k,x)$ are the
solutions to the full-line Schr\"odinger equation
with the respective asymptotic conditions
$$f_{\text{l}}(k,x)=e^{ikx}[1+o(1)],\quad
f'_{\text{l}}(k,x)=ik\,e^{ikx}[1+o(1)],\qquad x\to+\infty,$$
$$f_{\text{r}}(k,x)=e^{-ikx}[1+o(1)],\quad
f'_{\text{r}}(k,x)=-ik\,e^{-ikx}[1+o(1)],\qquad x\to-\infty.$$
In this case, $f_{\text{l}}(k,x)$
satisfies
$$f_{\text{l}}(k,x)=\ds\frac{e^{ikx}}{T(k)}+\ds\frac{L(k)\,e^{-ikx}}{T(k)},
\qquad x\le 0,\tag 5.12$$
and it agrees with [cf. (1.6)] the Jost solution $f(k,x)$ when $x\ge 0.$
Here, $L$ is the left reflection coefficient and
$T$ is the transmission coefficient. The right reflection
coefficient $R$ is given by
$$R(k)=-\ds\frac{L(-k)\,T(k)}{T(-k)},\qquad
k\in\bR.\tag 5.13$$
The potential can be uniquely reconstructed by using any one of the
full-line inversion methods [7,10,19-22] provided we can
construct the data
$\Cal F$ defined as
$$\Cal F:=\{L(k),T(k),R(k),\{\tau_j\}_{j=1}^N,\{c_{\text{l}j}\}_{j=1}^N,
\{c_{\text{r}j}\}_{j=1}^N,\{\gamma_j\}_{j=1}^N\},\tag 5.14$$
where the $-\tau_j^2$
correspond to the full-line bound-state energies. Note that
$T$ has poles at $k=i\tau_j$ in $\bCp$ for $j=1,\dots,N,$
the $c_{\text{l}j}$ are the norming constants defined as [cf. (3.24)]
$$c_{\text{l}j}:=\ds\frac{1}{\sqrt{\int_{-\infty}^\infty
dx\,f_{\text{l}}(i\tau_j,x)^2}},\qquad j=1,\dots,N,\tag 5.15$$
the $c_{\text{r}j}$ are the norming constants defined as in (5.15)
by replacing $f_{\text{l}}(k,x)$ with
$f_{\text{r}}(k,x),$ and the $\gamma_j$ are the bound-state
dependency constants defined as
$$\gamma_j:=\ds\frac{f_{\text{l}}(i\tau_j,x)}{f_{\text{r}}(i\tau_j,x)},
\qquad j=1,\dots,N.\tag 5.16$$
For example, in the Faddeev-Marchenko method [7,10,19-22] the potential $V$ and
$f_{\text{l}}(k,x)$
can be uniquely reconstructed as
$$V(x)=-2\,\ds\frac{dB_{\text{l}}(x,0^+)}{dx},\quad
f_{\text{l}}(k,x)=e^{ikx}\left[1+\int_0^\infty dy\,
B_{\text{l}}(x,y)\,e^{iky}\right],$$
where $B_{\text{l}}(x,y)$ is obtained by solving the left
Faddeev-Marchenko integral equation
$$B_{\text{l}}(x,y)+\Omega_{\text{l}}(2x+y)+\int_0^\infty dy\,
\Omega_{\text{l}}(2x+y+z)\,B_{\text{l}}(x,z)=0,\qquad y>0,$$
with the input data
$$\Omega_{\text{l}}(y):=\ds\frac{1}{2\pi}\int_{-\infty}^\infty
dk\,R(k)\,e^{iky}+\ds\sum_{j=1}^N c_{\text{l}j}^2\,
e^{-\tau_j y}.$$
Equivalently,
the potential $V$ and $f_{\text{r}}(k,x)$
can be uniquely reconstructed as
$$V(x)=2\,\ds\frac{dB_{\text{r}}(x,0^+)}{dx},\quad
f_{\text{r}}(k,x)=e^{-ikx}\left[1+\int_0^\infty dy\,
B_{\text{r}}(x,y)\,e^{iky}\right],$$
where $B_{\text{r}}(x,y)$ is obtained by solving the right
Faddeev-Marchenko integral equation
$$B_{\text{r}}(x,y)+\Omega_{\text{r}}(-2x+y)+\int_0^\infty dy\,
\Omega_{\text{r}}(-2x+y+z)\,B_{\text{r}}(x,z)=0,\qquad y>0,$$
with the input data
$$\Omega_{\text{r}}(y):=\ds\frac{1}{2\pi}\int_{-\infty}^\infty
dk\,L(k)\,e^{iky}+\ds\sum_{j=1}^N c_{\text{r}j}^2\,
e^{-\tau_j y}.$$

Let us now describe the construction of $\Cal F$ given in (5.14)
 from $\{F_\alpha,
F_\beta,\alpha,\beta\}$ with $\alpha\ne\beta$ or from $\{F_\pi,
F_\beta,\beta\}$ with $\beta\ne\pi,$ enabling us to use any of the
full-line inversion methods to reconstruct $V.$
Using (5.12) and its $x$-derivative evaluated at $x=0,$ we get
$$L(k)=\ds\frac{ik\,f(k,0)-f'(k,0)}{ik\,f(k,0)+f'(k,0)},
\quad T(k)=\ds\frac{2ik}{ik\,f(k,0)+f'(k,0)}.\tag 5.17$$
If $\alpha\ne \beta,$ with the
help of (2.2), (2.3), and (5.17), for $k\in\bCpb$ we obtain
$$L(k)=\cases
\ds\frac{(k-i\cot\beta)\,F_\alpha(k)-
(k-i\cot\alpha)\,F_\beta(k)}{(k+i\cot\beta)\,F_\alpha(k)-
(k+i\cot\alpha)\,F_\beta(k)},\qquad \alpha,\beta\in(0,\pi),\\
\stretch
\ds\frac{(k-i\cot\beta)\,F_\pi(k)-
F_\beta(k)}{(k+i\cot\beta)\,F_\pi(k)+
F_\beta(k)},\qquad \beta\in(0,\pi),\endcases\tag 5.18$$
$$T(k)=\cases
\ds\frac{2ikh_{\beta\alpha}}{
(k+i\cot\beta)\,F_\alpha(k)-
(k+i\cot\alpha)\,F_\beta(k)},\qquad \alpha,\beta\in(0,\pi),\\
\stretch
\ds\frac{2k}{
(k+i\cot\beta)\,F_\pi(k)+
F_\beta(k)},\qquad \beta\in(0,\pi)
,\endcases\tag 5.19$$
and using (5.13), for $k\in\bR$ we get
$$R(k)=\cases\ds\frac{-(k+i\cot\beta)\,F_\alpha(-k)+
(k+i\cot\alpha)\,F_\beta(-k)}{
(k+i\cot\beta)\,F_\alpha(k)-
(k+i\cot\alpha)\,F_\beta(k)},\qquad \qquad \alpha,\beta\in(0,\pi),\\
\stretch
-\ds\frac{(k+i\cot\beta)\,F_\pi(-k)+F_\beta(-k)}
{(k+i\cot\beta)\,F_\pi(k)+F_\beta(k)},\qquad \beta\in(0,\pi).\endcases\tag 5.20$$
Since $V\equiv 0$ for $x<0,$ it is already known
that the norming constants
$c_{\text{r}j}$ are related [13,21] to the residues of
$L$ at the poles $k=i\tau_j$ as
$$c_{\text{r}j}=\ds\sqrt{-i\,\text{Res}(L,i\tau_j)},\qquad
j=1,\dots,N.\tag 5.21$$
Using (5.12) and the fact that
$f_{\text{r}}(k,x)=e^{-ikx}$ for $x\le 0,$ we have
$$\gamma_j=f_{\text l}(i\tau_j,0)=f(i\tau_j,0)=\ds\frac{L}{T}(i\tau_j)=
\ds\frac{\text{Res}(L,i\tau_j)}{\text{Res}(T,i\tau_j)},\tag 5.22$$
and then via (5.15) and (5.16) we get
$$c_{\text{l}j}=\ds\frac{c_{\text{r}j}}{|\gamma_j|}=
\ds\frac{(-1)^{N-j}c_{\text{r}j}}{\gamma_j}=
\ds\frac{i(-1)^{N-j+1}\text{Res}(T,i\tau_j)}{
\sqrt{-i\,\text{Res}(L,i\tau_j)}},\tag 5.23$$
where we have used the fact [21] that
the sign of $\gamma_j$ is the same as that of
$(-1)^{N-j}.$

\vskip 10 pt
\noindent {\bf 6. EXAMPLES}
\vskip 3 pt

In this section, we illustrate the uniqueness and recovery
described in Theorems~2.1-2.8 with some concrete examples. The
existence of a potential in the Faddeev class corresponding to the
scattering data in each example is assured by verifying that the
corresponding left reflection coefficient $L$ satisfies the
characterization conditions given in Theorem~3.3 of [39]. In these
examples, the Jost functions and scattering coefficients are rational functions
of $k;$ consequently, the integral equations of Gel'fand-Levitan,
Marchenko, and Faddeev-Marchenko have degenerate kernels,
enabling us to solve them explicitly and to recover the related
potentials in closed forms. Such potentials are known as Bargmann
potentials and they decay exponentially as $x\to+\infty.$

\noindent {\bf Example~6.1} In the data $\Cal D_1$ of Theorem~2.1,
let us specify
$$N_\alpha=0,\quad N_\beta=0,\quad
|F_\alpha(k)|^2=k^2+c^2
\ \text{ for \ } k\in\bR,$$
for some fixed $c,$
but let us leave the value of $h_{\beta\alpha}$ as yet an
unspecified parameter. Since $F_\alpha(0)\ne 0,$
we cannot have $c=0$ and hence
we can assume $c>0.$ Proceeding as in the proof of
Theorem~2.1, we obtain
$$F_\alpha(k)=k+ic,\quad
\text{Re}[\Lambda_1(k)]=\ds\frac{h_{\beta\alpha}k}{k^2+c^2},\quad
\Lambda_1(k)=\ds\frac{h_{\beta\alpha}}{k+ic}.$$
Using (4.3) we get $\cot\alpha=-c,$
and hence via (2.1) we have
$\cot\beta=h_{\beta\alpha}-c.$
We also obtain $F_\beta(k)=k+i(c-h_{\beta\alpha}).$
Since $N_\beta=0$ we must have $h_{\beta\alpha}\le c.$
Then, from (2.2) and (2.3) we get
$f(k,0)=1$ and $f'(k,0)=ik.$ Thus, $V=0$ is the unique
potential corresponding to the data, regardless of
the value of $h_{\beta\alpha}.$ Unless $h_{\beta\alpha}$
is specified, we cannot determine $\beta$ and $F_\beta$
uniquely as they both contain the parameter
$h_{\beta\alpha}.$

\noindent {\bf Example~6.2} In the data $\Cal D_2$ of Theorem~2.2,
let us specify
$$N_\pi=0,\quad N_\beta=0,\quad
|F_\pi(k)|^2=1
\ \text{ for \ } k\in\bR,$$
but let us leave the value of $\beta$ as yet an
unspecified parameter. Proceeding as in the proof of
Theorem~2.2, we find that $V(x)=0$ is the unique
potential corresponding to our data, regardless of
the value of $\beta.$
We get
$$F_\pi(k)=1,\quad
\text{Re}[\Lambda_2(k)]=0,\quad
\Lambda_2(k)=0,\quad F_\beta(k)=k-i\cot\beta,$$
with the only restriction on $\beta$ given by
$\beta\in[\pi/2,\pi),$ or equivalently $\cot\beta\le 0,$ so that
$N_\beta=0.$ Thus, unless the value
of $\beta$ is specified in $\Cal D_2,$ we cannot
uniquely determine $F_\beta.$

\noindent {\bf Example~6.3} In the data $\Cal D_1$ of Theorem~2.1,
let us specify
$$N_\alpha=1,\quad N_\beta=1,\quad
\kappa_{\alpha 1}=2,
\quad \kappa_{\beta 1}=4,\quad
|F_\alpha(k)|^2=\ds\frac{(k^2+1)(k^2+4)}{k^2+16}
\ \text{ for \ } k\in\bR,$$
but let us leave the value of $h_{\beta\alpha}$ as yet an
unspecified parameter. Proceeding as in the proof of
Theorem~2.1, we find
$$F_\alpha(k)=\ds\frac{(k+i)(k-2i)}{k+4i},\quad
\text{Re}[\Lambda_1(k)]=\ds\frac{h_{\beta\alpha}k}{k^2+1},\quad
\Lambda_1(k)=\ds\frac{h_{\beta\alpha}}{k+i}.$$
Then, as
$k\to\infty$ we get $\Lambda_1(k)=\ds\frac{h_{\beta\alpha}}{k}-
\ds\frac{ih_{\beta\alpha}}{k^2}+O(1/k^3),$ and a comparison with
(4.3) indicates that
$\cot\alpha=-1+\ds\frac{12}{h_{\beta\alpha}}.$ Next, with the
help of (2.1) we get
$\cot\beta=h_{\beta\alpha}-1+\ds\frac{12}{h_{\beta\alpha}}.$ We also get
$$F_\beta(k)=\ds\frac{(k-4i)[k+i(1-h_{\beta\alpha})]}{k+2i}.$$
Note
that we must have $h_{\beta\alpha}\le 1$ in order to ensure that
$N_\beta=1;$ thus, we get the restriction
$h_{\beta\alpha}\in(0,1].$ Via (2.2) we obtain
$$f(k,0)=\ds\frac{k^2+12ik/h_{\beta\alpha}+
(16-12/h_{\beta\alpha})}{(k+2i)(k+4i)}.$$
A straightforward analysis indicates that $f(k,0)$ has no zeros in
$\bCp$ if $h_{\beta\alpha}\in(0,3/4),$ the two zeros of $f(k,0)$
are $k=0$ and $k=-16i$ if $h_{\beta\alpha}=3/4,$ and $f(k,0)$ has
exactly one zero in $\bCp$ if $h_{\beta\alpha}\in(3/4,1].$ Unless
the value of $h_{\beta\alpha}$ is specified in the data, we get a
one-parameter family for each of $V,$ $\alpha,$ and $\beta,$ where
$h_{\beta\alpha}$ is the parameter.
Using (5.18) and (5.19) we obtain
$$L(k)=\ds\frac{6}{\xi(k,h_{\beta\alpha})}\,
\left[k(1-2/h_{\beta\alpha}+12/h_{\beta\alpha}^2)
+i(1-17/h_{\beta\alpha}+12/h_{\beta\alpha}^2)\right],$$
$$T(k)=\ds\frac{1}{\xi(k,h_{\beta\alpha})}\,
\left[k(k+2i)(k+4i)\right],$$
where
$$\xi(k,h_{\beta\alpha}):=k^3+
12ik^2/h_{\beta\alpha}+(10-72/h_{\beta\alpha}^2)k
+(-6+102/h_{\beta\alpha}-72/h_{\beta\alpha}^2)i.$$
The corresponding one-parameter family
of potentials can be obtained by using any of the methods outlined in Section~5.

\noindent {\bf Example~6.4} As the data $\Cal D_2$ of Theorem~2.2,
let us specify
$$N_\pi=1,\quad N_\beta=1,\quad \kappa_{\pi 1}=2,\quad
\kappa_{\beta 1}=4,\quad
|F_\pi(k)|^2=\ds\frac{k^2+4}{k^2+20} \
\text{ for \ } k\in\bR,$$
but let us leave the value of $\beta$ as yet an
unspecified parameter. Proceeding as in the proof of Theorem~2.2, we obtain
$$F_\pi(k)=\ds\frac{k-2i}{k+\sqrt{20}\,i},\quad
\text{Re}[\Lambda_2(k)]=\ds\frac{4}{k^2+16},\quad
\Lambda_2(k)=\ds\frac{i}{k+4i},$$
$$F_\beta(k)=\ds\frac{(k-4i)[k^2+ik(4-\cot\beta)+4(1+\cot\beta)]}
{(k+2i)(k+\sqrt{20}\,i)}.$$
We find
that $F_\beta$ has no zeros in $\bCp$ other than $k=4i$ if
$\cot\beta<-1;$ its zeros are $k=0,$ $k=-5i,$ and $k=4i$ if
$\cot\beta=-1;$ and it has a second zero in $\bCp$ other than
$k=4i$ if $\cot\beta>-1.$
Thus, for consonance with $N_\beta=1,$ we must have
$\beta\in[\cot^{-1}(-1),\pi).$ Unless the value of $\beta$ is
specified, we get a one-parameter family of potentials for the given $\Cal D_2.$
The corresponding scattering coefficients are obtained via (5.18) and (5.19) as
$$L(k)=\ds\frac{-8k+(8+6\cot\beta)i}
{k^3+12k-(8+6\cot\beta)i},\quad
T(k)=\ds\frac{k(k+2i)(k+\sqrt{20}i)}{k^3+12k-(8+6\cot\beta)i}.$$

\noindent {\bf Example~6.5} In the data $\Cal D_3$ of Theorem~2.3,
let us specify
$$N_\alpha=0,\quad N_\beta=1,\quad
|F_\alpha(k)|^2=k^2+4 \
\text{  for \ }
k\in\bR,$$
but let us leave the value
of $h_{\beta\alpha}$ as yet an unspecified parameter. Proceeding
as in the proof of Theorem~2.3, we get $F_\alpha(k)=k+2i.$
Using (4.14) we find
$$\Cal H(k,\kappa)=\ds\frac{ih_{\beta\alpha}k}{(\kappa+2)(k+i\kappa)(k+2i)},$$
and hence $\ds\lim_{k\to \infty} [k\,\Cal
H(k,\kappa)]=\ds\frac{ih_{\beta\alpha}}{\kappa+2}.$ The value of
$\kappa_{\beta 1}$ is then obtained via (4.15) as $\kappa_{\beta
1}=h_{\beta\alpha}-2.$ Note that we must have $h_{\beta\alpha}>2$
because $\kappa_{\beta 1}$ must be positive. We also get
$$F_\beta(k)=k-i(h_{\beta\alpha}-2),\quad
\Lambda_3(k)=\ds\frac{ik}{(k+2i)[k+i(h_{\beta\alpha}-2)]},$$
and hence as $k\to\infty$ we obtain
$$\Lambda_3(k)=\ds\frac{i}{k}+\ds\frac{h_{\beta\alpha}}
{k^2}+\ds\frac{i[-2h_{\beta\alpha}-(h_{\beta\alpha}-2)^2]}{k^3}+O(1/k^4).$$
Thus, from (4.11) we find
$\cot\alpha=-2,$ and then via (2.1) we get
$\cot\beta=h_{\beta\alpha}+2.$ Hence, unless the value of
$h_{\beta\alpha}$ is specified in $\Cal D_3,$ we find a
one-parameter family for each of $\beta$ and $F_\beta.$
On the other hand, from (2.2) we get $f(k,0)=1$ and
$V$ is uniquely determined as
$V(x)=0$. With the help
(5.18) and (5.19), we get
$L(k)=0$ and $T(k)=1,$ respectively.

\noindent {\bf Example~6.6} In the data $\Cal D_3$ of Theorem~2.3,
let us specify
$$N_\alpha=1,\quad N_\beta=2,\quad \kappa_{\alpha 1}=2,\quad
\kappa_{\beta 2}=4,\quad
|F_\alpha(k)|^2=k^2+4 \
\text{ for \ } k\in\bR,$$
but let us leave the value of $h_{\beta\alpha}$ as yet an
unspecified parameter. We get $F_\alpha(k)=k-2i.$ From the interlacing property stated in
Proposition~3.7, we have the restriction $\kappa_{\beta 1}\in(0,2).$
Proceeding as in the proof of Theorem~2.3, we find
$$\Cal H(k,\kappa)=\ds\frac{ih_{\beta\alpha}k}{(\kappa+4)(k+i\kappa)(k+4i)},\quad
\ds\lim_{k\to \infty} [k\,\Cal
H(k,\kappa)]=\ds\frac{ih_{\beta\alpha}}{\kappa+4}.$$
The value of
$\kappa_{\beta 1}$ is then obtained via (4.15) as $\kappa_{\beta
1}=h_{\beta\alpha}-4.$ Thus, the restriction $\kappa_{\beta 1}\in
(0,2)$ indicates that $h_{\beta\alpha}\in(4,6).$ We also get
$$\Lambda_3(k)=\ds\frac{ik}{(k+4i)[k+i(h_{\beta\alpha}-4)]},
\quad F_\beta(k)=\ds\frac{(k-4i)[k-i(h_{\beta\alpha}-4)]}{k+2i}.$$
Using (4.11) and then (2.1) we obtain
$$\cot\alpha=\ds\frac{12}{h_{\beta\alpha}}-4,\quad
\cot\beta=h+\ds\frac{12}{h_{\beta\alpha}}-4.$$
Via (2.2) we get
$f(k,0)=\ds\frac{k-i(4-12/h_{\beta\alpha})}{k+2i},$ and we find
that $f(k,0)$ has exactly one zero in $\bCp$ when
$h_{\beta\alpha}\in(4,6).$ With the help of (5.18) and (5.19),
we have
$$L(k)=\ds\frac{6(h_{\beta\alpha}-6)(h_{\beta\alpha}-2)/h_{\beta\alpha}^2}{\eta(k,h_{\beta\alpha})},
\quad
T(k)=\ds\frac{k(k+2i)}{\eta(k,h_{\beta\alpha})},$$
where
$$\eta(k,h_{\beta\alpha}):=k^2+
(-4+12/h_{\beta\alpha})ik+(-6+48/h_{\beta\alpha}-72/h_{\beta\alpha}^2).$$
Note that unless
the value of $h_{\beta\alpha}$ is specified in $\Cal D_3,$ we find
a one-parameter family for each of $V,$ $\alpha,$ and $\beta.$
The corresponding one-parameter family
of potentials can be obtained by using any of the methods outlined in Section~5.

\noindent {\bf Example~6.7} In the data $\Cal D_4$ of Theorem~2.4,
let us specify
$$N_\pi=1,\quad
N_\beta=2,\quad
\kappa_{\pi 1}=2,
\quad
\kappa_{\beta 1}=1,\quad
|F_\pi(k)|^2=\ds\frac{k^2+4}{k^2+1}\
\text{ for } k\in\bR,$$
but let us
leave the value of $\beta$ as yet an unspecified parameter.
Proceeding as in the proof of Theorem~2.4, we find
$$F_\pi(k)=\ds\frac{k-2i}{k+i},\quad
\Cal H(k,\kappa)=\ds\frac{-i\kappa}{k+i\kappa}.$$
With the help of (4.21) we get
$$\kappa_{\beta 2}=\cot\beta,\quad
F_\beta(k)=\ds\frac{(k-i)(k-i\cot\beta)}
{k+2i}.$$
Using the second line of (5.18) and of (5.19) we obtain
$$L(k)=-\ds\frac{3i\cot\beta}{2k^3+5k+3i\cot\beta},\quad
T(k)=\ds\frac{2k(k+i)(k+2i)}{2k^3+5k+3i\cot\beta}.$$
When $\beta$ is specified, the unique potential corresponding to our data
can be obtained by using any of the methods specified in Section~5.
On the other hand, if
the value of $\beta$ is left unspecified in $\Cal D_4,$ we find
a one-parameter family for each of $V$ and $F_\beta.$

\noindent {\bf Example~6.8} In the data $\Cal D_5$ of Theorem~2.5,
let us specify
$$N_\alpha=1,\quad N_\beta=1,\quad \kappa_{\alpha 1}=2,\quad
\kappa_{\beta 1}=4,\quad
|F_\beta(k)|^2=\ds\frac{(k^2+1)(k^2+16)}{k^2+4}\
\text{ for \ } k\in\bR,$$
but let us leave the value of $h_{\beta\alpha}$ as yet an
unspecified parameter. Proceeding as in the proof of
Theorem~2.4, we find
$$F_\beta(k)=\ds\frac{(k+i)(k-4i)}{k+2i},\quad
\text{Re}[\Lambda_5(k)]=\ds\frac{-h_{\beta\alpha}}{k^2+1},\quad
\Lambda_5(k)=\ds\frac{-ih_{\beta\alpha}}{k+i}.$$
Using (4.25) and then (2.1) we
obtain
$$\cot\beta=-1+12/h_{\beta\alpha},\quad
\cot\alpha=-1-h_{\beta\alpha}+12/h_{\beta\alpha}.$$
We also find
$F_\alpha(k)=\ds\frac{(k-2i)[k+i(1+h_{\beta\alpha})]}{k+4i}.$ We
check that $F_\alpha$ has exactly one zero in $\bCp$ because
$h_{\beta\alpha}>0.$ With the help of (2.2) we get
$$f(k,0)=\ds\frac{k^2+12ik/h_{\beta\alpha}+(4-12/h_{\beta\alpha})}{(k+2i)(k+4i)},$$
and find that $N_\pi=1$ if $h_{\beta\alpha}>3,$
$N_\pi=0$ if $h_{\beta\alpha}\in(0,3),$ and the two zeros of $f(k,0)$ when
$h_{\beta\alpha}=3$ are given by $k=0$ and $k=-4i.$
Using (5.18) and (5.19) we get
$$L(k)=\ds\frac{-6k(1+2/h_{\beta\alpha}-12/h^2_{\beta\alpha})
-6i(1+5/h_{\beta\alpha}-12/h^2_{\beta\alpha})}
{\zeta(k,h_{\beta\alpha})},\quad
T(k)=\ds\frac{k(k+2i)(k+4i)}{\zeta(k,h_{\beta\alpha})},$$
where
$$\zeta(k,h_{\beta\alpha}):=k^3+
12ik^2/h_{\beta\alpha}+(10-72/h_{\beta\alpha}^2)k+(6+30/h_{\beta\alpha}-72/h_{\beta\alpha}^2)i.$$
Unless the
value of $h_{\beta\alpha}$ is specified in the data, we get a
one-parameter family for each of $V,$ $\alpha,$ and $\beta.$
The corresponding potentials can be obtained
by using one of the methods outlined in Section~5.

\noindent {\bf Example~6.9} In the data $\Cal D_6$ of Theorem~2.6,
let us specify
$$N_\pi=1,\quad N_\beta=1,\quad \kappa_{\pi 1}=1,\quad
\kappa_{\beta 1}=2,\quad
|F_\beta(k)|^2=k^2+4\
\text{ for \ } k\in\bR.$$
Proceeding as in the proof of Theorem~2.6, we find
$$F_\beta(k)=k-2i,\quad
\text{Re}[\Lambda_6(k)]=\ds\frac{-1}{k^2+1},\quad
\Lambda_6(k)=\ds\frac{-i}{k+i},$$
$$F_\pi(k)=\ds\frac{k-i}{k+2i},\quad
\cot\beta=-1,\quad
f'(k,0)=\ds\frac{i(k+i)(k-2i)}{k+2i}.$$
Our data $\Cal D_6$
uniquely determines $V$ via
the Gel'fand-Levitan method or the Marchenko method. As
outlined at the end of Section~5, the same
potential can also be obtained by any of the full-line inversion
methods by evaluating the full-line reflection and
transmission coefficients, which are obtained via (5.18) and (5.19), respectively, as
$$L(k)=\ds\frac{-3/2}{k^2-ik+3/2},\quad
T(k)=\ds\frac{k(k+2i)}{k^2-ik+3/2}.$$
 From the pole of $T$ in $\bCp$
we see that the full-line problem has one bound state at
$k=i(\sqrt{7}+1)/2.$ The norming constant in the left Faddeev-Marchenko
data is given by $c_{{\text l}1}=\sqrt{6+33\sqrt{7}/14}.$

\noindent {\bf Example~6.10} In our data $\Cal D_7$ in Theorem~2.7,
let us specify
$$N_\alpha=0,\quad N_\beta=1,\quad \kappa_{\beta 1}=2,
\quad |F_\beta(k)|^2=k^2+4\
\text{ for \ } k\in\bR,$$
but let us leave the values
of $\beta$ and $h_{\beta\alpha}$ as yet unspecified parameters.
Proceeding as in the proof of Theorem~2.7, we
obtain
$$F_\beta(k)=k-2i,\quad
\text{Re}[\Lambda_7(k)]=0,\quad\Lambda_7(k)=0,$$
$$P(k)=4-h_{\beta\alpha}\cot\beta,\quad
F_\alpha(k)=\ds\frac{k^2+ih_{\beta\alpha}k+(4-h_{\beta\alpha}\cot\beta)}
{k+2i}.$$
Because we assume $N_\alpha=0,$ none of the two zeros
of $F_\alpha$ are allowed to be in $\bCp,$ and hence
we must have $h_{\beta\alpha}\cot\beta\ge 4.$ We find that
$f(k,0)=\ds\frac{k+i\cot\beta}{k+2i},$ and hence $V$ can be
specified uniquely if and only if the value of $\beta$ is
specified in the data. Otherwise, there is a one-parameter family
of potentials.
Via (2.1) we
have $\cot\alpha=\cot\beta-h_{\beta\alpha};$ hence, leaving both
$\cot\beta$ and $h_{\beta\alpha}$ unspecified in our data results
in a two-parameter family for each of $\alpha$ and $F_\alpha.$
We also get
$$L(k)=\ds\frac{-4+\cot^2\beta}{2k^2+2ik\,\cot\beta+(4-\cot^2\beta)},\quad
T(k)=\ds\frac{2k(k+2i)}{2k^2+2ik\,\cot\beta+(4-\cot^2\beta)}.\tag 6.1$$
Using any one of the recovery methods outlined in Section~5, we obtain
$$V(x)=\ds\frac{32(\cot^2\beta-4)\,e^{-4x}}
{\left[(\cot\beta+2)-(\cot\beta-2)e^{-4x}\right]^2},\qquad \cot\beta\ge 2.$$
When $\cot\beta\in[4/h_{\beta\alpha},2),$ the expression for $V$ can be obtained
explicitly, but it is too long to display here. It has the form
$$V(x)=-\ds\frac{d}{dx}\left[\ds\frac{\Gamma'(x,b)}
{\Gamma(x,b)}\right],\qquad \cot\beta\in[0,2),$$
with
$$\Gamma(x,b):=c_1+c_2e^{-4x}+c_3e^{-(\sqrt{8-\cot^2\beta}-\cot\beta)x}
+c_4e^{-(4+\sqrt{8-\cot^2\beta}-\cot\beta)x},$$
where the coefficients $c_1,$ $c_2,$ $c_3,$ and $c_4$ are
independent of $x$ and they are certain explicit functions of $\cot\beta.$

\noindent {\bf Example~6.11} In the data $\Cal D_7$ of Theorem~2.7,
let us specify
$N_\alpha=1,$ $N_\beta=2,$ and
$$\kappa_{\alpha 1}=2,
\quad\kappa_{\beta 1}=1,\quad\kappa_{\beta 2}=4,
\quad |F_\beta(k)|^2=\ds\frac{(k^2+1)(k^2+16)}{k^2+4}
\ \text{ for \  } k\in\bR,$$
but let us leave the values of $\beta$ and $h_{\beta\alpha}$ as
yet unspecified parameters. Proceeding as in the proof of
Theorem~2.7, we find
$$F_\beta(k)=\ds\frac{(k-i)(k-4i)}{k+2i},
\quad\text{Re}[\Lambda_7(k)]=0,\quad
\Lambda_7(k)=0.$$
 From (4.31) we get
$P(k)=13-h_{\beta\alpha}\cot\beta,$ and we find
$$F_\alpha(k)=\ds\frac{(k-2i)[k^2+
ih_{\beta\alpha}k+(13-h_{\beta\alpha}\cot\beta)} {(k+i)(k+4i)}.$$
Because we assume $N_\alpha=1,$ $F_\alpha$ must not have any
zeros in $\bCp$ other than $k=2i,$ which is the case if ${h_{\beta\alpha}}\cot\beta\ge
13.$ We also find that
$$f(k,0)=\ds\frac{k^3+ik^2\cot\beta+
4k+i(4\cot\beta-36/h_{\beta\alpha})}{(k+i)(k+2i)(k+4i)},$$
$$L(k)=\ds\frac{h_{\beta\alpha}(\cot^2\beta-13)k^2
-36ik+(4h_{\beta\alpha}\cot^2\beta -16h_{\beta\alpha}-36\cot\beta)}
{\omega(k,h_{\beta\alpha},\beta)},$$
$$T(k)=\ds\frac{2h_{\beta\alpha}k(k+i)
(k+2i)(k+4i)}{\omega(k,h_{\beta\alpha},\beta)},$$
where we have defined
$$\aligned
\omega(k,h_{\beta\alpha},\beta):=&
2h_{\beta\alpha}k^4+2ih_{\beta\alpha}k^3\cot\beta
+h_{\beta\alpha}(21-\cot^2\beta)
k^2\\
& +(8h_{\beta\alpha}\cot\beta-36)ik+(36\cot\beta+
16h_{\beta\alpha}-4h_{\beta\alpha}\cot^2\beta).
\endaligned$$
Hence, $V$ can be specified uniquely if and only if both $\beta$
and $h_{\beta\alpha}$ are specified in $\Cal D_7;$
otherwise, a two-parameter family of
potentials corresponds to $\Cal D_7.$

\noindent {\bf Example~6.12} In the data $\Cal D_8$ of Theorem~2.8,
let us specify
$$N_\pi=0,\quad N_\beta=1,\quad \kappa_{\beta 1}=2,\quad
|F_\beta(k)|^2=k^2+4\ \text{ for \ } k\in\bR,$$
but let us leave the
value of $\beta$ as yet an unspecified parameter. Proceeding
as in the proof of Theorem~2.8, we find
$$F_\beta(k)=k-2i,\quad \text{Re}[\Lambda_8(k)]=0,\quad
\Lambda_8(k)=0,\quad F_\pi(k)=\ds\frac{k+i\cot\beta}{k+2i}.$$
We also get the same $L$ and $T$ given in (6.1).
In order to have
$N_\pi=0,$ we must have $\cot\beta\ge 0,$ or equivalently
$\beta\in(0,\pi/2].$ Because $N_\pi=0,$
the potential $V$ is uniquely determined
 from $F_\pi$ if the value of $\beta$ is specified
in our data; otherwise, we get a one-parameter
family of potentials that are described in Example~6.10.

\noindent {\bf Example~6.13} In the data $\Cal D_8$ of Theorem~2.8,
let us specify $N_\pi=1,$ $N_\beta=2,$ and
$$\kappa_{\alpha 1}=2,\quad
\kappa_{\beta 1}=1,\quad \kappa_{\beta 2}=4,\quad
|F_\beta(k)|^2=\ds\frac{(k^2+1)(k^2+16)}{k^2+4}
\ \text{ for \ } k\in\bR,$$
but let us leave the value of $\beta$ as yet an unspecified
parameter. Proceeding as in the proof of Theorem~2.8, we find
$$F_\beta(k)=\ds\frac{(k-i)(k-4i)}{k+2i},\quad
\text{Re}[\Lambda_8(k)]=0,\quad \Lambda_8(k)=0,\quad
F_\pi(k)=\ds\frac{(k-2i)(k+i\cot\beta)}{(k+i)(k+4i)}.$$
We also find
$$L(k)=\ds\frac{(\cot^2\beta-13)k^2+(4\cot^2\beta-16)}{\chi(k,\cot\beta)},\quad
T(k)=\ds\frac{2k(k+2i)(k+4i)}{\chi(k,\cot\beta)},$$
where
$$\chi(k,\cot\beta):=2k^4+2ik^3\cot\beta+(21-\cot^2\beta)
k^2+8ik\cot\beta+(16-4\cot^2\beta).$$
Once the value of $\beta$ is specified
in our data, the potential $V$ can uniquely be
determined by using one of the recovery methods outlined
in Section~5; otherwise, we get a one-parameter
family of potentials depending on the parameter
$\beta.$

\vskip 10 pt

\noindent {\bf Acknowledgment.} The research leading to this
article was supported in part by the National Science Foundation
under grant DMS-0204437, the Department of Energy under grant
DE-FG02-01ER45951, and Universidad Nacional Aut\'onoma de M\'exico
under Proyecto PAPIIT-DGAPA IN 101902.

\vskip 10 pt

\noindent {\bf{References}}

\vskip 3 pt

\item{[1]} V. A. Ambartsumyan, {\it \"Uber eine Frage der
Eigenwerttheorie,} Z. Phys. {\bf 53}, 690--695 (1929).

\item{[2]} G. Borg, {\it Eine Umkehrung der Sturm-Liouvilleschen
Eigenwertaufgabe,} Acta Math. {\bf 78}, 1--96 (1946).

\item{[3]} G. Borg, {\it Uniqueness theorems in the spectral
theory of $y''+(\lambda-q(x))y=0,$} Proc. 11th Scandinavian
Congress of Mathematicians, Johan Grundt Tanums Forlag, Oslo,
1952, pp. 276--287.

\item{[4]} V. A. Marchenko, {\it Some questions in the theory of
one-dimensional linear differential operators of the second order.
I}, Trudy Moskov. Mat. Ob\v s\v c. {\bf 1}, 327--420
(1952) [Am. Math. Soc. Transl. (ser. 2) {\bf 101}, 1--104 (1973)].

\item{[5]} F. Gesztesy and B. Simon, {\it Uniqueness theorems in
inverse spectral theory for one-dimensional Schr\"odinger
operators,} Transac. Am. Math. Soc. {\bf 348}, 349--373 (1996).

\item{[6]} B. M. Levitan and M. G. Gasymov, {\it Determination of
a differential equation by two of its spectra,} Uspekhi Mat. Nauk {\bf 19},
3--63 (1964) [Russian Math.
Surveys {\bf 19}, 1--63 (1964)].

\item{[7]}  V.\; A.\; Marchenko, {\it Sturm-Liouville operators and
applications,} Birk\-h\"au\-ser, Basel, 1986.

\item{[8]} B. M. Levitan , {\it Inverse Sturm-Liouville problems,}
VNU Science Press, Utrecht, 1987.

\item{[9]} J. Weidmann, {\it Spectral theory of ordinary
differential operators,} Lecture Notes in Math. {\bf 1258},
Springer, Berlin, 1987.

\item{[10]} K. Chadan and P. C. Sabatier, {\it Inverse problems in quantum
scattering theory,} 2nd ed., Springer, New York, 1989.

\item{[11]} I. M. Gel'fand and B. M. Levitan,
{\it On the determination of a differential equation from its
spectral function,} Izvestiya Akad. Nauk SSSR. Ser. Mat. {\bf 15}, 309--360
(1951) [Am. Math. Soc. Transl. (ser. 2) {\bf 1}, 253--304 (1955)].

\item{[12]} V. A. Marchenko,
{\it On reconstruction of the potential energy from phases of the scattered waves,}
Dokl. Akad. Nauk SSSR {\bf 104}, 695--698 (1955).

\item{[13]} N. N. Novikova and V. M. Markushevich,
{\it Uniqueness of the solution of the one-dimensional problem of
scattering for potentials located on the positive semiaxis,}
Vychisl. Seysmol. {\bf 18}, 176--184 (1985) [Comput. Seismology
{\bf 18}, 164--172 (1987)].

\item{[14]} T. Aktosun, M. Klaus, and C. van der Mee C,
{\it On the Riemann-Hilbert problem for the one-dimensional
Schr\"odinger equation,} J. Math. Phys. {\bf 34}, 2651--2690
(1993).

\item{[15]} T. Aktosun,
{\it Bound states and inverse scattering for the Schr\"odinger
equation in one dimension,} J. Math. Phys. {\bf 35}, 6231--6236
(1994).

\item{[16]} B. Gr\'ebert and R. Weder,
{\it Reconstruction of a potential on the line that is a priori
known on the half line,} SIAM J. Appl. Math. {\bf 55}, 242--254
(1995).

\item{[17]} T. Aktosun,
{\it Inverse Schr\"odinger scattering on the line with partial
knowledge of the potential,} SIAM J. Appl. Math. {\bf 56},
219--231 (1996).

\item{[18]} F. Gesztesy and B. Simon,
{\it Inverse spectral analysis with partial information on the
potential. I. The case of an a.c. component in the spectrum,}
Helv. Phys. Acta {\bf 70}, 66--71 (1997).

\item{[19]} L. D. Faddeev, {\it Properties of the $S$-matrix of the one-dimensional
Schr\"odinger equation,} Trudy Mat. Inst. Steklov. {\bf 73}, 314--336
(1964) [Am. Math. Soc. Transl. (ser. 2) {\bf 65}, 139--166 (1967)].

\item{[20]} A. Melin,
{\it Operator methods for inverse scattering on the real line,}
Comm. Partial Differential Equations {\bf 10}, 677--766 (1985).

\item{[21]} T. Aktosun and M. Klaus, {\it
Chapter 2.2.4, Inverse theory: problem on the line,} In: E. R.
Pike and P. C. Sabatier (eds.), {\it Scattering,} Academic Press,
London, 2001, pp. 770--785.

\item{[22]} P. Deift and E. Trubowitz, {\it
Inverse scattering on the line,} Comm. Pure Appl. Math. {\bf 32},
121--251 (1979).

\item{[23]} T. Aktosun,
{\it Construction of the half-line potential from the Jost
function,} IMA preprint \#1926, May 2003.

\item{[24]} B. J. Forbes, E. R. Pike, and D. B. Sharp,
{\it The acoustical Klein-Gordon equation: The wave-mechanical
step and barrier potential functions,} J. Acoust. Soc. Am. {\bf
114}, 1291--1302 (2003).

\item{[25]} E. C. Titchmarsh,
{\it Introduction to the theory of Fourier integral,} 2nd ed.,
Oxford University Press, London, 1962.

\item{[26]} L. Ahlfors, {\it Complex analysis,}
2nd ed., McGraw-Hill, New York, 1966.

\item{[27]} R. V. Churchill and J. W. Brown, {\it Complex
variables and applications,}4th ed., McGraw-Hill, New York, 1984.

\item{[28]} L. D. Faddeev, {\it
The inverse problem in the quantum theory of scattering,} Uspekhi Mat. Nauk
{\bf 14}, 57--119 (1959) [J. Math.
Phys. {\bf 4}, 72--104 (1963)].

\item{[29]} R. G. Newton, {\it Scattering theory of waves and particles,}
2nd ed., Springer, New York, 1982.

\item{[30]} K. Chadan, {\it Chapter 2.2.1,
Potential scattering: the radial Schr\"odinger equation,} In: E.
R. Pike and P. C. Sabatier (eds.), {\it Scattering,} Academic
Press, London, 2001, pp. 669--685.

\item{[31]} K. Chadan and P. C. Sabatier, {\it Chapter 2.2.1,
Radial inverse scattering problems,} In: E.
R. Pike and P. C. Sabatier (eds.), {\it Scattering,} Academic
Press, London, 2001, pp. 726--741.

\item{[32]} T. Aktosun and M. Klaus, {\it
Small-energy asymptotics for the Schr\"odinger equation on the
line,} Inverse Problems {\bf 17}, 619--632 (2001).

\item{[33]} M. Reed and B. Simon, {\it Methods of modern
mathematical physics, II, Fourier analysis, self-adjointness,}
Academic Press, New York, 1975.

\item{[34]} T. Kato, {\it Perturbation theory for linear
operators,} 2nd ed., Springer, Berlin, 1976.

\item{[35]} R. A. Adams, {\it Sobolev spaces,} Academic Press, New York, 1975.

\item{[36]} E. A. Coddington and N. Levinson, {\it Theory of ordinary differential equations,}
McGraw-Hill, New York, 1955.

\item{[37]} E. C. Titchmarsh, {\it Eigenfunction expansions
associated with second-order differential equations, Part I,}
2nd ed., Clarendon Press, Oxford, 1962.

\item{[38]} B. M. Levitan and I. S. Sargsjan, {\it Introduction to spectral theory: selfadjoint
ordinary differential operators,} Am. Math. Soc., Providence, 1975.

\item{[39]} T. Aktosun and V. G. Papanicolaou,
{\it Recovery of a potential from the ratio of reflection and transmission coefficients,}
J. Math. Phys. {\bf 44}, 4875--4883 (2003).

\end